# Generalized Madelung transformations for quantum wave equations I: generalized spherical coordinates for field spaces

D. H. Delphenich

Lindsborg, KS USA 67456
e-mail: david.delphenich@yahoo.com

The Madelung transformation of the space in which a quantum wave function takes its values is generalized from complex numbers to include field spaces that contain orbits of groups that are diffeomorphic to spheres. The general form for the resulting real wave equations then involves structure constants for the matrix algebra that is associated with the group action. The particular cases of the algebras of complex numbers, quaternions, and complex quaternions, which pertain to the Klein-Gordon equation, the relativistic Pauli equation, and the bi-Dirac equation, resp., are then discussed.

## 1 Introduction

The two fundamental problems at the interface between physics and mathematics are the problem of the mathematical modeling of physical phenomena and the converse problem of the interpretation of mathematical results in terms of physical phenomena. Although the former problem is the one that defines the mathematical methods of physics most essentially, nevertheless, in the last century the latter problem has also been increasingly dominant, especially in the realm of quantum phenomena, which, by their nature, lie beyond the scope of human intuition or direct observation.

The problem of interpretation is complicated by the fact that the same mathematical models generally describe many analogous, but nonetheless distinct, physical phenomena. For instance, the same mathematical methods, namely, the calculus of exterior differential forms, can be applied with only revisions of the notations to the description of Maxwell's theory of electromagnetism, relativistic hydrodynamics, and the description of weak-field gravity, when one includes the "gravito-magnetic" contribution to the gravitational field. Hence, unless one knows in advance the physical context of a given mathematical model, choosing a proper physical interpretation for a mathematical model is a much deeper problem than it sounds like it should be on the surface of things.

If one examines the early history of quantum physics, one sees that this was precisely the nature of the greatest problem that early quantum physicists were confronting. In effect, the mathematical models were preceding the physical interpretations, so there was considerable debate about what the proper interpretations should be.

The dominant interpretation of wave mechanics that emerged was the statistical interpretation that was developed by the Copenhagen school of Bohr, Born, Heisenberg, and others. Basically, this stemmed from regarding the modulus squared $\rho = \psi^* \psi$ of the quantum wave function $\psi$ as a probability density function for the position of a point particle. However, not everyone agreed that the statistical interpretation represented a definitive theoretical statement. Notably, Einstein, who was himself distinguished in statistical physics, felt that the nature of statistical methodology was too inherently empirical to represent a fundamental statement of natural law

Another interpretation that came out in roughly the same year as the statistical interpretation was the hydrodynamical interpretation of Madelung [**1**], who suggested that instead of interpreting $\rho$ as a probability density function for a point particle, one should regard it as proportional to the mass density function of an extended matter distribution. The Schrödinger equation in the complex wave function $\psi$ could then be converted into a pair of real equations that took the form of the conservation of mass for $\rho$ and the conservation of energy for the extended distribution. In the latter equation, the distinction between classical and quantum continuum mechanics took the form of a "quantum potential" function that was proportional to $\Delta\sqrt{\rho}/\sqrt{\rho}$, by way of Planck's constant. Hence, the problem of interpretation was reduced to the problem of correctly accounting for the appearance of the quantum potential.

The general nature of Madelung transform is that one attempts to convert a wave equation in a complex-valued wave function into a pair of real equations by means of expressing the values of the wave function in polar coordinates in the usual method of setting $\psi = Re^{i\theta}$. Although the extension of the Madelung transformation from the Schrödinger equation to the Klein-Gordon equation was carried out by Takabayasi [**2**], with analogous results, nonetheless, when later researchers attempted to give a "neo-classical" form to the Pauli equation [**3**] and the Dirac equation [**4**], the path that they chose was the related, but not identical, method of bilinear covariants, which is the generally accepted method that is taught in modern wave mechanics. (For excellent surveys on the causal, or "hydrodynamical," interpretation of wave mechanics, see Vigier [**5**], de Broglie [**6**], and Halbwachs [**7**].)

It is in the context of the Klein-Gordon equation that yet another interpretation for $\rho$ emerged, namely the Pauli-Weisskopf interpretation. Since the original reasons for rejecting the Klein-Gordon equation as the physically correct relativistic form of the Schrödinger equation included the fact that the conserved current that one derives from its Lagrangian is not positive definite, and thus cannot be consistent with the nature of probability currents, it is interesting that some time later Pauli and Weisskopf [**8**] decided that the Klein-Gordon equation was a reasonable candidate for a relativistic wave equation if one chose to interpret $\rho$ as an electric charge density.

Now, an interesting aspect of the wave functions for the Pauli equation, which is the Schrödinger equation when the particle has spin, and the Dirac equation is that they take their values in vector spaces, namely, $\mathbb{C}^2$ and $\mathbb{C}^4$, which take the form of the product manifolds $\mathbb{R}^+ \times S^3$ and $\mathbb{C}^* \times S^3_{\mathbb{C}}$ when one ignores the origin. (Here, our notation is that $\mathbb{R}^+$ denotes the positive real numbers, $S^3$ denotes the (real) 3-sphere, $\mathbb{C}^*$ denotes the non-zero complex numbers, and $S^3_{\mathbb{C}}$ refers to the complex 3-sphere.) If one considers that $\mathbb{C}^*$ takes the form of $\mathbb{R}^+ \times S^1$ then one sees that all of the physically interesting wave functions – at least at the elementary level – seem to take their values in field spaces that take the form of a radial coordinate and a sphere.

Furthermore, the spheres in question are also the underlying manifolds of Lie groups in each case. In fact, if one regards the motion of the wave as involving not only a group of motions that acts on the spacetime manifold, but another – possibly the same – group that acts on the space of field values then it is natural to transfer the values of the wave functions to the group of motions that acts on field space.

The purpose of the present article is then to generalize the Madelung transformation by generalizing the spherical coordinates introduced on the field space. The resulting system of real equations is then derived in general form and then specialized to each of the cases of interest to wave mechanics, namely, $\mathbb{C}$, $\mathbb{C}^2$ and $\mathbb{C}^4$.

Since the resulting system of real differential equations involves the products of the matrices that generate the field space motions, not the Lie brackets, it is necessary to summarize some elementary notions from the theory of associative algebras, which is presented in an Appendix. The specific results for the algebras in question, namely, $\mathbb{C}$, the quaternions $\mathbb{H}$, and the complex quaternions $\mathbb{H}^{\mathbb{C}}$, are derived in section 2. In section 3, various forms for wave equations are presented, the basic idea of generalized spherical coordinates on the field spaces is defined, and the general form is obtained for the system of real equations that is associated with essentially the Klein-Gordon equation when one allows the wave function to take its values in a Lie group of the type in question. In section 4, this general form is specialized to each of the algebras under discussion.

This article stops short, however, of proposing a new interpretation for wave mechanics as a result of the analysis. This is because the analysis of the wave equations is incomplete, at this point, in advance of any discussion of the role of spin in the waves being described, along with the relationship between the present method and the method of bilinear covariants. These topics will then be addressed in subsequent efforts, since the presentation of the generalization of the Madelung transformation is already going to occupy a considerable volume of space.

## 2 Examples of associative algebras

We first consider how the abstract concepts that are discussed in the Appendix apply to in the specific cases of interest to quantum wave equations.

### 2.1 The complex numbers

The algebra $\mathbb{C}$, given the usual complex multiplication, can be regarded as an associative, commutative algebra with unity over $\mathbb{R}^2$; indeed, it is the complexification of the algebra $\mathbb{R}$.

We shall regard the set $\{1, i\} \subset \mathbb{R}^2$ as a real basis for $\mathbb{C}$, where 1 refers to the point $(1, 0) \in \mathbb{R}^2$ and $i$ to the point $(0, 1)$. The set $\{i\}$ represents a set of generators for the algebra $\mathbb{C}$ since $i^2 = -1$.

The representation of $\mathbb{C}$ in $M(2, \mathbb{R})$ that we shall use is defined by the left (or right) multiplication of complex numbers on $\mathbb{R}^2$. One need only map the basis $\{1, i\}$ to $\{I, J\} \subset M(2, \mathbb{R})$, resp., where:

$$I = \begin{bmatrix} 1 & 0 \\ 0 & 1 \end{bmatrix}, \qquad J = \begin{bmatrix} 0 & -1 \\ 1 & 0 \end{bmatrix}. \tag{2.1}$$

Hence, the complex number $a + ib$ will go to:

$$aI + bJ = \begin{bmatrix} a & -b \\ b & a \end{bmatrix}. \tag{2.2}$$

This representation is faithful, and the image of $\mathbb{R}^2$ in $M(2, \mathbb{R})$ is then a two-dimensional real vector subspace. Note that since $\mathbb{C}$ is generated by $i$, it is sufficient to define the matrix that represents $i$ in order to define any representation of $\mathbb{C}$.

If we introduce the notation $e_0 = I$, $e_1 = J$ then the algebra multiplication is defined by the relations:

$$(e_0)^2 = -(e_1)^2 = e_0, \qquad e_0\, e_1 = e_0 e_1 = e_1. \tag{2.3}$$

Since the algebra $\mathbb{C}$ is commutative, we have:

$$\{e_a, e_b\} = 2(\eta_{ab}\, e_0 + \alpha_{ab}\, e_1), \qquad [e_a, e_b] = 0 \quad (a, b = 0, 1), \tag{2.4}$$

in which the matrices $\eta_{ab}$ and $\alpha_{ab}$ are defined by:

$$\eta_{ab} = \begin{bmatrix} 1 & 0 \\ 0 & -1 \end{bmatrix}, \qquad \alpha_{ab} = \begin{bmatrix} 0 & 1 \\ 1 & 0 \end{bmatrix}, \tag{2.5}$$

which are seen to be the matrix of the two-dimensional Minkowski scalar product in an orthonormal frame and the matrix of the permutation of the $x$ and $y$ axis, or, what amounts to the same thing, the real and the imaginary numbers.

Hence, we can summarize the multiplication table for the algebra of $\mathbb{C}$ in the form:

$$e_a\, e_b = \eta_{ab}\, e_0 + \alpha_{ab}\, e_1 \qquad (a, b = 0, 1). \tag{2.6}$$

This allows us summarize the structure constants of $\mathbb{C}$ as:

$$a_{ab}^0 = 2b_{ab}^0 = \eta_{ab}, \qquad a_{ab}^1 = 2b_{ab}^1 = \alpha_{ab}, \qquad c_{ab}^c = 0. \tag{2.7}$$

The exponential map for the algebra $\mathbb{C}$ is the usual one that is given in complex analysis. Because the algebra is commutative, one can say that $\exp(a + b) = \exp(a)\exp(b)$ for any $a, b \in \mathbb{C}$, and, in particular:

$$\exp(a + ib) = \exp(a)\exp(ib). \tag{2.8}$$

The image of $\mathbb{C}$ under exp is then $\mathbb{C}^*$, which denotes the non-zero complex numbers. Hence, all numbers $\exp(a + ib)$ will have multiplicative inverses and, indeed, define the Abelian Lie group of non-zero complex numbers under multiplication. When one chooses an appropriate interval on the imaginary line, the map exp is also invertible, and the previous expression for exp(*a* + *ib*) represents the polar coordinate form for any non-zero complex number *z*, namely:

$$z = Re^{i\theta}. \tag{2.9}$$

One can then say that the multiplicative group $\mathbb{C}^* = \exp(\mathbb{C})$ is isomorphic to the direct product group $\mathbb{R}^+ \times U(1)$, where $\mathbb{R}^+$ represents the multiplicative group of positive real numbers.

If one regards $z = a + ib$ as a complex number, and chooses $\{1, i\}$ as a basis for $\mathbb{C}$ as a real two-dimensional algebra then the expansion of exp *z* into a linear combination of basis elements is given simply by de Moivre's theorem:

$$\exp z = e^a \cos b + i\, e^a \sin b. \tag{2.10}$$

The corresponding expression in terms of 2×2 real matrices when $z = aI + bJ$ is then:

$$\exp z = (e^a \cos b)\, I + (e^a \sin b)\, J. \tag{2.11}$$

2.2 The quaternions

We now extend the vector space $\mathbb{C}$ to two complex dimensions, which then represents four real dimensions. However, although the real vector space $\mathbb{R}^4$ can be regarded as the complexification of the real vector space $\mathbb{C} = \mathbb{R}^2$, nonetheless, when one examines the structure constants of the algebra $\mathbb{H}$ as an algebra over $\mathbb{R}^4$, one will see that it is *not*, in fact, the complexification of the *algebra* $\mathbb{C}$. In fact, we shall see that the subalgebra of $M(2; \mathbb{C})$ that we shall be ultimately concerned with is essentially real, not complex.

We denote the canonical basis elements of $\mathbb{R}^4$ by 1, **i**, **j**, **k**, respectively, and give them the multiplication table:

$$1^2 = -\mathbf{i}^2 = -\mathbf{j}^2 = -\mathbf{k}^2 = 1, \quad 1(\text{anything}) = (\text{anything})1 = \text{anything},$$
$$\mathbf{ij} = -\mathbf{ji} = \mathbf{k}, \quad \mathbf{jk} = -\mathbf{kj} = \mathbf{i}, \quad \mathbf{ki} = -\mathbf{ik} = \mathbf{j}.$$

We then recognize that we are dealing with the algebra of the *quaternions* (see, e.g., Porteus [9] or Artin [10]). It is associative, but from the last three sets of relations, one can see that it is not commutative.

Furthermore, since the squares of **i** and **j** are proportional to 1, while **ij** gives the basis element **k**, it is clear that the set {**i**, **j**} represents a minimal set of generators for the algebra. One can represent the basis set in the form {−**ii**, **i**, **j**, **ij**}.

We see that the subalgebra spanned by {1, **i**} is isomorphic to the algebra $\mathbb{C}$, but although we have extended the vector space from two real dimensions to four, nevertheless, we cannot regard $\mathbb{H}$ as the complexification of the algebra $\mathbb{C}$, nor any other algebra over $\mathbb{C}^2$. The key to understanding this is to recall (cf., Appendix) that the product map for a complex algebra must be $\mathbb{C}$-bilinear.

Hence, suppose we represent the vector space of $\mathbb{H}$ as $\mathbb{C} \oplus \mathbb{C}\mathbf{j}$. Let $u = a + b\mathbf{j}$ and $v = c + d\mathbf{j}$, with $a, b, c, d \in \mathbb{C}$. If the algebra product were $\mathbb{C}$-bilinear then one would have:

$$uv = (a + b\mathbf{j})(c + d\mathbf{j}) = (ac - bd) + (ac + bd)\mathbf{j} = vu. \qquad (2.12)$$

However, as we saw from the multiplication rules for the basis elements, the algebra is not commutative, in general.

The representation that we shall use in $M(2; \mathbb{C})$ is the real subalgebra that is spanned by the matrices $\{e_a, a = 0, \ldots, 3\}$ with:

$$e_0 = I, \qquad e_1 = J, \qquad e_2 = i\begin{bmatrix} 1 & 0 \\ 0 & -1 \end{bmatrix}, \qquad e_3 = i\begin{bmatrix} 0 & 1 \\ 1 & 0 \end{bmatrix}. \qquad (2.13)$$

The matrices $e_i$, $i = 1, 2, 3$ are proportional to the Pauli matrices $\sigma_i$ by way of the factor $i$, since we are using *skew*-Hermitian matrices, instead of Hermitian ones. This is because it is really the skew-Hermitian matrices that define the infinitesimal generators of one-parameter subgroups of unitary transformations, not the Hermitian ones. Of course, given a Hermitian matrix $a$ one can associate it with a skew-Hermitian one in the form of $ia$, so the distinction is mostly one of convenience. This is essentially a higher-dimensional analogue of regarding an element of $U(1)$ as $e^\theta$ with $\theta$ imaginary or as $e^{i\theta}$ with $\theta$ real. We choose the present convention since we shall be using both Hermitian and skew-Hermitian matrices in the next section.

We have also made a cyclic permutation of the basis elements [1], 123 → 231, in order to make the basis for $\mathbb{H}$ be an extension of the basis for $\mathbb{C}$. That is, in order relate our $e$ matrices to the Pauli $\sigma$ matrices, one must substitute:

$$e_1 = i\sigma_2, \qquad e_2 = i\sigma_3, \qquad e_3 = i\sigma_1. \qquad (2.14)$$

The multiplication table for the $e_a$ basis is obtained by direct computation:

$$(e_0)^2 = -(e_1)^2 = -(e_2)^2 = -(e_3)^2 = -e_0,$$

---

[1] Sakurai [**11**] refers to this particular permutation of the Pauli matrices as the "non-conformist" representation. However, the author of the present work is choosing it for the sake of mathematical convenience, not adolescent iconoclasm.

$$e_0 \, e_a = e_a e_0 = e_a, \quad \text{(all } a\text{)}$$
$$e_i \, e_j = \varepsilon_{ijk} \, e_k \quad (i, j = 1, 2, 3).$$

which makes the representation faithful, as one observes. For this algebra, it is sufficient to define the matrices that represent **i** and **j** in order to completely define the representation.

One can then see that the matrix that corresponds to an arbitrary element $(x^0, \ldots, x^3)$ in $\mathbb{R}^4$ for this choice of basis will take the form:

$$x^a e_a = \begin{bmatrix} x^0 + ix^2 & -x^1 + ix^3 \\ x^1 + ix^3 & x^0 - ix^2 \end{bmatrix} \equiv \begin{bmatrix} z^1 & -\overline{z}^2 \\ z^2 & \overline{z}^1 \end{bmatrix}. \tag{2.15}$$

The latter expression shows that although one can define an invertible map that takes $(u, v) \in \mathbb{C}^2$ to a matrix in $M(2; \mathbb{C})$ that is in the image of $\mathbb{H}$, nevertheless, the map is not complex linear. In particular, under complexification the quaternion $u + v\mathbf{j}$ corresponds to the 2×2 complex matrix:

$$uI + vJ = \begin{bmatrix} u & -v \\ v & u \end{bmatrix} = \begin{bmatrix} x^0 + ix^2 & -x^1 - ix^3 \\ -x^1 - ix^3 & x^0 + ix^2 \end{bmatrix}, \tag{2.16}$$

if we also set $u = x^0 + ix^2$ and $v = x^1 + ix^3$. Hence, this representation of the vector space $\mathbb{H} = \mathbb{C}^2$ in $M(2; \mathbb{C})$ differs fundamentally from the one described in (2.15).

From the multiplication table, we see that:

$$\{e_0, e_a\} = 2\, e_a, \qquad \{e_i, e_j\} = -2\delta_{ij}\, e_0, \tag{2.17a}$$
$$[e_0, e_a] = 0, \qquad [e_i, e_j] = 2\varepsilon_{ijk}\, e_k, \tag{2.17b}$$

which allows us to summarize the multiplication table as:

$$e_0\, e_a = 2\, e_a, \quad e_i\, e_j = -\delta_{ij}\, e_0 + \varepsilon_{ijk}\, e_k. \tag{2.18}$$

The non-zero structure constants can then be summarized as:

$$a_{0b}^c = 1/2\, b_{0b}^c = \delta_b^c, \qquad a_{ij}^0 = 1/2\, b_{ij}^0 = -\delta_{ij}, \qquad a_{ij}^k = 1/2\, c_{ij}^k = \varepsilon_{ijk}. \tag{2.19}$$

Since the matrix $e_0$ commutes with everything, we can say that:

$$\exp(\phi^\mu e_\mu) = \exp(\phi^0 e_0 + \phi^i e_i) = \exp(\phi^0 e_0) \exp(\phi^i e_i) \equiv R\, e^{\phi^i e_i}. \tag{2.20}$$

The latter notation is appropriate since $\exp(\phi^0 e_0)$ will be a positive real number (times the identity matrix) and $\exp(\phi^i e_i)$ will be an element of $SU(2)$. That is, the group that we are dealing with is $\mathbb{R}^+ \times SU(2)$.

Since the squares of the matrices $e_i$ are proportional to $I = e_0$ any polynomial in them – in particular, $\exp(\phi^i e_i)$ – will have to be expanded into four components relative to the full set of basis elements of $\mathbb{H}$, not merely three components relative to the set of three that are defined by $e_i$. In other words, the vector space that is spanned by the set $\{e_i, i = 1, 2, 3\}$ does not define a subalgebra of $\mathbb{H}$ when one restricts the product. Hence, in general, one will have:

$$\Phi = \exp(\phi^i e_i) = \Phi^0 I + \Phi^i e_i . \tag{2.21}$$

For instance, if one is considering a rotation of angle $\theta$ about only one of the axes – say, $i = m$ – then one reverts to the previous result for the algebra $\mathbb{C}$, namely:

$$\exp(\theta e_m) = (\cos\theta) I + (\sin\theta) e_m . \tag{2.22}$$

### 2.3 The complex quaternions

As we pointed out in the Appendix, the algebra $\mathbb{H}$ is a non-relativistic version of a larger relativistic picture. We shall see that what is missing is contained in the complexification of $\mathbb{H}$, which we denote by $\mathbb{H}^\mathbb{C}$. However, we will also see that there are things that do not have analogues in Minkowski space.

The key to understanding how relativistic mechanics is the complexification of non-relativistic mechanics is to see that the identity component $SO_0(3,1)$ of the Lorentz group, which is sometimes called the *proper isochronous Lorentz group*, is isomorphic to $SO(3; \mathbb{C})$, as well as to $SL(2; \mathbb{C})$. We shall now examine both of these isomorphisms, although it will be necessary in what follows to consider only the infinitesimal versions that pertain to the Lie algebras, where one finds that the Lie algebra $\mathfrak{so}(3, 1)$ is isomorphic to the complexification of $SO(3; \mathbb{R})$.

The Lie algebra $\mathfrak{so}(3, 1)$ of the Lorentz group can be defined by a subalgebra of the Lie algebra that is associated with the associative algebra $M(4; \mathbb{R})$. Specifically, it is defined by the vector subspace that consists of all 4×4 real matrices $a \in M(4; \mathbb{R})$ such that:

$$a + a^* = 0, \tag{2.23}$$

where the asterisk refers to the *Lorentz adjoint* of $a$, which we define by:

$$a^* = \eta a^T \eta, \tag{2.24}$$

in which $\eta = \text{diag}[+1, -1, -1, -1]$ and the "T" signifies the matrix transpose. Hence, one can also represent the defining relation as:

$$\eta a + a^T \eta = 0, \tag{2.25}$$

which is the infinitesimal version of:

$$A^T \eta A = \eta, \tag{2.26}$$

which defines the matrix $A$ to be a finite Lorentz transformation.

The Lie algebra $\mathfrak{so}(3, 1)$ is six-dimensional as a real vector space and one can define a basis for it by means of the six matrices:

$$J_1 = \begin{bmatrix} 0 & 0 & 0 & 0 \\ 0 & 0 & 0 & 0 \\ 0 & 0 & 0 & -1 \\ 0 & 0 & 1 & 0 \end{bmatrix}, \quad J_2 = \begin{bmatrix} 0 & 0 & 0 & 0 \\ 0 & 0 & 0 & 1 \\ 0 & 0 & 0 & 0 \\ 0 & -1 & 0 & 0 \end{bmatrix}, \quad J_3 = \begin{bmatrix} 0 & 0 & 0 & 0 \\ 0 & 0 & -1 & 0 \\ 0 & 1 & 0 & 0 \\ 0 & 0 & 0 & 0 \end{bmatrix}, \tag{2.27a}$$

$$K_1 = \begin{bmatrix} 0 & 1 & 0 & 0 \\ 1 & 0 & 0 & 0 \\ 0 & 0 & 0 & 0 \\ 0 & 0 & 0 & 0 \end{bmatrix}, \quad K_2 = \begin{bmatrix} 0 & 0 & 1 & 0 \\ 0 & 0 & 0 & 0 \\ 1 & 0 & 0 & 0 \\ 0 & 0 & 0 & 0 \end{bmatrix}, \quad K_3 = \begin{bmatrix} 0 & 0 & 0 & 1 \\ 0 & 0 & 0 & 0 \\ 0 & 0 & 0 & 0 \\ 1 & 0 & 0 & 0 \end{bmatrix}. \tag{2.27b}$$

The matrices $J_i$, $i = 1, 2, 3$ then represent the infinitesimal generators of Euclidian rotations along the $x$, $y$, and $z$ axes, resp., while the matrices $J_i$, $i = 1, 2, 3$ represent the infinitesimal generators of Lorentz boosts along those axes.

The commutation relations for these basis elements then take form:

$$[J_i, J_j] = 2\varepsilon_{ijk} J_k, \qquad [J_i, K_j] = 2\varepsilon_{ijk} K_k, \qquad [K_i, K_j] = -2\varepsilon_{ijk} J_k. \tag{2.28}$$

The first set defines the Lie algebra of $\mathfrak{so}(3; \mathbb{R})$. The second set says that the adjoint action of $\mathfrak{so}(3; \mathbb{R})$ on the vector space of infinitesimal boosts takes boosts to other boosts. The last set is particularly interesting from the standpoint of Thomas precession since it leads to a purely relativistic rotation of a spin vector when it is not in a state of rectilinear motion.

Now, observe what happens when one allows the $J_i$ basis vectors to be multiplied by complex scalars in such a way that:

$$K_i = iJ_i. \tag{2.29}$$

These three vectors then span a three-dimensional complex vector space over the same set of vectors as the six-dimensional real one that $\mathfrak{so}(3, 1)$ is defined over. In fact, one can associate each 4×4 real matrix in $\mathfrak{so}(3, 1)$ that takes the form $\omega^i J_i + \beta^i K_i$ with a complex 3×3 matrix of the form $(\omega^i + i\beta^i)J_i$, with:

$$J_1 = \begin{bmatrix} 0 & 0 & 0 \\ 0 & 0 & -1 \\ 0 & 1 & 0 \end{bmatrix}, \qquad J_2 = \begin{bmatrix} 0 & 0 & 1 \\ 0 & 0 & 0 \\ -1 & 0 & 0 \end{bmatrix}, \qquad J_3 = \begin{bmatrix} 0 & -1 & 0 \\ 1 & 0 & 0 \\ 0 & 0 & 0 \end{bmatrix}, \qquad (2.30)$$

this time. These are clearly the infinitesimal generators of three-dimensional Euclidian rotations about the *x*, *y*, and *z* axes, respectively, although now we are also allowing them to be multiplied by complex scalars.

Furthermore, if one makes these replacements in the commutation rules, one sees that the second and third sets take the form:

$$[J_i, iJ_j] = 2\varepsilon_{ijk}\, iJ_k\,, \qquad [iJ_i, iJ_j] = -2\varepsilon_{ijk}\, J_k\,. \qquad (2.31)$$

Now, the Lie bracket that is associated with $M(3; \mathbb{C})$ is complex bilinear, so these commutation relations are consistent with the assumption that we are dealing with the complex Lie algebra spanned by the three $J_i$ with the commutation relations of $\mathfrak{so}(3)$. Hence, the assignment of $(\omega^i + i\beta^i)J_i$ with $\omega^i J_i + \beta^i K_i$ says that the complexification of the Lie algebra $\mathfrak{so}(3; \mathbb{R})$ is (real) isomorphic to the Lie algebra $\mathfrak{so}(3, 1)$.

It is interesting to note that since the vector cross product on $\mathbb{R}^3$ gives it the structure of the Lie algebra $\mathfrak{so}(3; \mathbb{R})$, all that one has to do to make the vector cross product useful in relativistic mechanics is to define it over $\mathbb{C}^3$, instead of $\mathbb{R}^3$, and in the same manner. Hence, one does not necessarily have to abandon the cross product when going from non-relativistic to relativistic physics, merely complexify it.

Now, let us look at the Lie algebra $\mathfrak{sl}(2; \mathbb{C})$, which is also isomorphic to $\mathfrak{so}(3, 1)$. It consists of 2×2 complex matrices with trace zero, so it is a three-complex-dimensional complex vector subspace of $M(2; \mathbb{C})$, as well as a Lie subalgebra under the associated Lie bracket. One can then regard it as a six-dimensional real vector subspace. Indeed, we have already defined a three-real-dimensional subspace of $\mathfrak{sl}(2; \mathbb{C})$ in the form of $\mathfrak{su}(2)$, which is also isomorphic to the Lie algebra $\mathfrak{so}(3)$. Hence, all that one must to do to complete $\mathfrak{su}(2)$ to $\mathfrak{sl}(2; \mathbb{C})$ is complexify it.

As we pointed out above, when one multiplies a skew-Hermitian matrix by *i* one obtains a Hermitian one. Now, one can polarize any $a \in \mathfrak{sl}(2; \mathbb{C})$ with respect to the Hermitian conjugate operator:

$$a = \tfrac{1}{2}(a - a^\dagger) + \tfrac{1}{2}(a + a^\dagger) \equiv \omega + i\beta, \tag{2.32}$$

such that $\omega, \beta \in \mathfrak{su}(2)$.

Hence, all that is necessary in order to define the (complex) isomorphism of $\mathfrak{so}(3; \mathbb{C})$ with $\mathfrak{sl}(2; \mathbb{C})$ is to map the basis elements $J_i \in \mathfrak{so}(3; \mathbb{C})$ to the basis elements $e_i \in \mathfrak{su}(2)$ in the obvious way. The infinitesimal Euclidian rotations are then represented by 2×2 complex skew-Hermitian matrices and the infinitesimal boosts are represented by Hermitian ones. Note that although the Lie algebra $\mathfrak{su}(2)$ is represented, in this case, by complex matrices that act most naturally on the complex vector space $\mathbb{C}$, nevertheless, it is a *real* Lie algebra, and its complexification is then $\mathfrak{sl}(2; \mathbb{C})$.

Having said all of the foregoing about the complexification of $\mathfrak{so}(3; \mathbb{R})$ and $\mathfrak{su}(2)$, we now expand our discussion to the complexification of the algebra $\mathbb{H}$.

The complexification $\mathbb{H}^\mathbb{C}$ of $\mathbb{H}$ can be regarded as the algebra over $\mathbb{C}^4$ that one obtains by first starting with the (real) basis vectors $\{1, \mathbf{i}, \mathbf{j}, \mathbf{k}\}$ for $\mathbb{H}$ and multiplying them by complex scalar coefficients, as long as one clearly distinguishes the imaginary $i$ from the vector $\mathbf{i}$. Nonetheless, one still maintains the same multiplication table for the basis elements, with the understanding that the multiplication of elements must be complex bilinear now. Note that the center of the algebra $\mathbb{H}^\mathbb{C}$ − viz., all elements that commute with everything − is the complexification of the center of $\mathbb{H}$. That is, the center of $\mathbb{H}$ consists of all real scalar multiples of 1, whereas the center of $\mathbb{H}^\mathbb{C}$ consists of all complex scalar multiples of 1.

Here, we must point out that whereas the complexification of the spatial part of $\mathbb{H}$ has a corresponding Minkowski space interpretation in terms of infinitesimal rotations and infinitesimal boosts, nevertheless, only the real part of the center, which is isomorphic to $\mathbb{C}^*$, has any Minkowski space interpretation, namely, as a dilatation; the imaginary part of a complex dilatation seems to be something beyond the scope of Minkowski space.

To see what it represents, consider that even though $\mathbb{C}^4$ has a canonical decomposition as a real vector space $\mathbb{R}^4 \oplus i\mathbb{R}^4$ into a real and an imaginary subspace, nonetheless, if $V$ is a complex $n$-dimensional vector space and $V'$ is a real $n$-dimensional vector space then there is no unique way of expressing $V$ as $V' \oplus iV'$, unless, for example, one defines a basis for each vector space. Hence, there will be extra degrees of freedom associated with the choice of a "real + imaginary" splitting of $V$.

If the complex dilatation $\lambda$ takes the form $Re^{i\theta}$ then one sees that the effect of the $e^{i\theta}$ part on a given complex basis $\{\mathbf{e}_a, a = 1, \ldots, n\}$, or rather its associated real basis $\{\mathbf{e}_a, \bar{\mathbf{e}}_a,$ $a = 1, \ldots, n\}$ amounts to a collective rotation of each pair $\{\mathbf{e}_a, \overline{\mathbf{e}}_a\}$ in the plane that they span:

$$\mathbf{e}_a \to \cos\theta\, \mathbf{e}_a - \sin\theta\, \overline{\mathbf{e}}_a, \qquad \overline{\mathbf{e}}_a \to \sin\theta\, \mathbf{e}_a + \cos\theta\, \overline{\mathbf{e}}_a. \tag{2.33}$$

Thus, the $e^{i\theta}$ part of $\lambda$ represents a duality rotation for the expression of $\mathbb{C}^4$ in terms of $\mathbb{R}^4 \oplus i\mathbb{R}^4$.

Notice that the projection $\mathbb{C}^* \to \mathbb{R}^*$, $Re^{i\theta} \mapsto R$ defines a homomorphism of the multiplicative groups, so it is the $U(1)$ part of $\mathbb{C}^*$ that is basically lost under the projection.

Analogous to the way that we complexified $\mathbb{H}$ itself, we can also complexify the representation of $\mathbb{H}$ in $M(2; \mathbb{C})$ by means of the basis matrices $\{e_a, a = 0, 1, 2, 3\}$. Since these matrices span a four-dimensional real vector subspace, they will span a four-complex-dimensional subspace under complexification. However, $M(2; \mathbb{C})$ is four-complex-dimensional to begin with, so we are faithfully representing $\mathbb{H}^\mathbb{C}$ by way of the entire algebra of $M(2; \mathbb{C})$.

The missing four-real-dimensional subspace in $M(2; \mathbb{C})$ that is added during complexification is spanned by the $e_a$ when they are given imaginary coefficients, or rather, by the $ie_a$ when they are given real coefficients. Hence, $ie_0$ represents the imaginary axis in $\mathbb{C}$, while the $ie_i$, $i = 1, 2, 3$ are Hermitian matrices. Here, we see why it is necessary to be consistent in our use of anti-Hermitian matrices to represent infinitesimal rotations, instead of Hermitian ones, since we are also using the Hermitian matrices to represent other things.

As a complex algebra over $\mathbb{C}^4$, the multiplication table of $\mathbb{H}^\mathbb{C}$, and therefore the set of structure constants, is same as for $\mathbb{H}$, namely, (2.18), but its multiplication table as a real algebra over $\mathbb{R}^8$ must take into account the fact that $i^2 = -1$, since the real vector $a$ and the imaginary vector $ia$ are linearly independent over the real scalars, but not over the complex scalars. Hence, we extend our real basis $\{e_a, a = 0, \ldots, 3\}$ for $\mathbb{H}$ to a real basis $\{e_a, \overline{e}_a \equiv ie_a, a = 0, \ldots, 3\}$ for $\mathbb{H}^\mathbb{C}$ and include the extra entries in the multiplication table:

$$\overline{e}_a e_b = e_a \overline{e}_b = i\, e_a e_b = a_{ab}^c \overline{e}_c, \qquad \overline{e}_a \overline{e}_b = -e_a e_b = -a_{ab}^c e_c. \tag{2.34}$$

Therefore, we must extend the structure constants $a_{ab}^c$ for $\mathbb{H}$ accordingly:

$$a^{\bar{c}}_{ab} = a^{\bar{c}}_{\bar{a}b} = -a^{c}_{\bar{a}\bar{b}} = a^{c}_{ab}, \tag{2.35}$$

all other constants being null.

One can see from (2.34) that since the "imaginary" basis elements $\bar{e}_a$ produce "real" basis elements when one forms products of them, a minimal set of generators for the algebra $\mathbb{H}^\mathbb{C}$ can be defined by the set of elements $\{\bar{e}_a, a = 0, \ldots, 3\}$.

Since $e_0$ commutes with everything, under the exponential map, a general element $\phi^a e_a \in \mathbb{H}^\mathbb{C}$, where the $\phi^a$ are complex now, maps to:

$$\exp(\phi^a e_a) = \exp(\phi^0) \exp(\phi^i e_i) = (Re^{i\theta}) \exp(\phi^i e_i), \tag{2.36}$$

if we assume that $\exp(\phi^0)$ is represented in polar form by $Re^{i\theta}$.

Thus, the group that we are considering in this case is $\mathbb{C}^* \times SL(2; \mathbb{C})$, which is isomorphic to $GL(2; \mathbb{C})$. The isomorphism amounts to either multiplication of a matrix in $SL(2; \mathbb{C})$ by a non-zero complex scalar or the factorization of a general invertible 2×2 complex matrix $A \in GL(2; \mathbb{C})$ into a product $(\det A)^{1/2} [(\det A)^{-1/2} A]$ of a non-zero complex scalar and a unimodular matrix.

Since the complex dimension of the Lie algebra $\mathfrak{gl}(2; \mathbb{C})$ is four, we can no longer embed it in $\mathbb{C}^2$, as we did for $\mathbb{R} \oplus \mathfrak{su}(2)$, but must use $\mathbb{C}^4$ instead. Hence, we also need a faithful representation of the *real* algebra $\mathbb{H}^\mathbb{C}$ in $M(4; \mathbb{C})$. We take advantage of the fact that $\mathbb{H}^\mathbb{C} = \mathbb{H} \oplus i\mathbb{H}$ and first represent multiplication by 1 as multiplication by the identity matrix $I$, and multiplication by $i$ as multiplication by the 4×4 matrix:

$$J = \begin{bmatrix} 0 & -I \\ I & 0 \end{bmatrix}. \tag{2.37}$$

in which the $I$ matrices are 2×2 identity matrices.

We next represent each 2×2 matrix $e_a$ as the direct sum 4×4 matrix:

$$E_a = \begin{bmatrix} e_a & 0 \\ 0 & e_a \end{bmatrix}. \tag{2.39}$$

The matrices $\bar{e}_a = ie_a$ then correspond to:

$$\overline{E}_a = J E_a = \begin{bmatrix} 0 & -e_a \\ e_a & 0 \end{bmatrix}. \tag{2.40}$$

One sees that the matrices $\overline{E}_i$, $i = 1, 2, 3$ have essentially the same form as the Dirac $\alpha_i$ matrices:

$$\alpha_i = \begin{bmatrix} 0 & \sigma_i \\ \sigma_i & 0 \end{bmatrix}, \qquad i = \mu = 1, 2, 3. \tag{2.41}$$

when one makes the replacements listed in (2.13).

There is, however, a significant difference between $\overline{E}_0 = J$ and the Dirac $\beta$ matrix:

$$\beta = \begin{bmatrix} I & 0 \\ 0 & -I \end{bmatrix}, \tag{2.42}$$

namely, the $\beta$ matrix does not belong to the algebra that is spanned by the eight basis elements $E_a$ and $\overline{E}_a$, which must all take the form:

$$\phi^a E_a + \overline{\phi}^a \overline{E}_a = \begin{bmatrix} \phi^a e_a & -\overline{\phi}^a e_a \\ \overline{\phi}^a e_a & \phi^a e_a \end{bmatrix}, \qquad \phi^a, \overline{\phi}^a \in \mathbb{R}. \tag{2.43}$$

There is clearly no such linear combination that will result in the matrix $\beta$. Hence, the Clifford algebra generated by the set $\{\beta, \alpha_i, i = 1, 2, 3\}$ is quite distinct from the algebra $\mathbb{H}^{\mathbb{C}}$, even though the anti-commutators of the basis elements of $\mathbb{H}^{\mathbb{C}}$ seem to have much in common with those of the aforementioned Clifford algebra.

We can, however, say that:

$$\alpha_i = i\beta \overline{E}_i, \qquad i = 1, 2, 3 \tag{2.44}$$

up to a permutation of the spatial indices.

Recall that the Dirac $\gamma$ matrices are defined by [2]:

$$\gamma_0 = \beta, \qquad \gamma_i = \begin{bmatrix} 0 & \sigma_i \\ -\sigma_i & 0 \end{bmatrix} = \beta \alpha_i, \tag{2.45}$$

---

[2] Since various conventions exist for the gamma matrices, which generally depend upon the choice of sign convention for the Minkowski scalar product, we are presenting the Bjorken and Drell form [**12**], which is consistent with our choice of convention on $g$ that it be "timelike positive." The form presented in Sakurai [**11**], which differs by a factor of $i$ for the spatial indices, is based on representing the Minkowski scalar product with the "imaginary time" convention that makes it Euclidian.

so we can relate the spatial matrices to our corresponding basis elements by way of:

$$\gamma_i = i\,\overline{E}_i, \tag{2.46}$$

up to a permutation of the spatial indices.

One must note that since the vector space $M(4;\mathbb{C})$ has complex dimension 16 – hence, real dimension 32 – the complex representation of $\mathbb{H}^{\mathbb{C}}$ by the four $e$ matrices, or the real representation by the eight $e$ and $\overline{e}$ matrices, does not span the entire vector space of the representation. Similarly, the Clifford algebra that is generated by the four $\gamma$ matrices is defined over a vector space of real dimension 16, which does not exhaust the vector space $M(4;\mathbb{C})$, either.

## 3 Wave equations

The physics of waves is more general in scope than any particular wave equation, whether linear or nonlinear, and strictly begins in the definition of constitutive laws for the wave medium, along with field equations for the wave field. However, this generally leads to dispersion laws for the medium, which eventually come down to a Lorentzian structure $g$ (or possibly more than one) on the tangent or cotangent bundle of the configuration manifold $M$; for simplicity, we assume that $M$ is an open submanifold of $\mathbb{R}^n$, which amounts to a choice of coordinate system on $M$.

### 3.1 Forms for wave equations

Due to the nature of the dispersion laws for most wave media, most wave equations will generally involve the resulting d'Alembertian operator:

$$\Box \equiv g^{\mu\nu}\frac{\partial}{\partial x^\mu}\frac{\partial}{\partial x^\nu} \tag{3.1}$$

in some way.

In the quasi-linear case, the wave equation for $\psi$ will take the form:

$$\Box\,\psi = F(x^\mu, \psi, \psi_{,\mu}). \tag{3.2}$$

In the linear case, these equations often take the form of the eigenvalue equation:

$$\Box\,\psi = \lambda\psi, \tag{3.3}$$

in which the wave function $\psi: M \to V$ takes its values in a vector space $V$, which might be real or complex, and $\lambda \in \mathbb{C}$ is a scalar *constant*.

One can put the general second order quasilinear equation into first-order form by essentially the "1-jet prolongation" of $\psi$, namely, the addition of its first partial derivatives as supplementary variables:

$$\psi_{,\mu} = k_\mu, \qquad g^{\mu\nu} k_{\mu,\nu} = F(x^\mu, \psi, k_\mu). \tag{3.4}$$

If the components $g^{\mu\nu}$ are constant on $M$ then one can put this system of equations into "conservation law" form:

$$\psi_{,\mu} = k_\mu, \qquad k^\mu{}_{,\mu} = F(x^\mu, \psi, k_\mu), \qquad k^\mu = g^{\mu\nu} k_\nu. \tag{3.5}$$

In fact, if the components of $g$ are not constant in this coordinate system then it is more correct to make the left-hand side of the middle equation in (3.5) take the form:

$$k^\mu{}_{;\mu} = \frac{1}{\sqrt{|g|}} \nabla_\mu (\sqrt{|g|}\, k^\mu), \tag{3.6}$$

in which the covariant derivative operator $\nabla$ relates to the Levi-Civita connection of $g$.

One can further simplify the form of (3.5) by introducing the calculus of exterior differential forms. Namely, if one sets $k = k_\mu\, dx^\mu$, which then makes $k$ a 1-form on $M$ with values in $V$, then (3.5) can be expressed in the coordinate-free form:

$$d\psi = k, \qquad \delta \mathbf{k} = F(j^1 \psi), \qquad \mathbf{k} = i_g k. \tag{3.7}$$

The notation used here is: $d$ refers to the exterior derivative operator, which acts on exterior differential forms, $\delta$ to the divergence operator, which acts on multivector fields, $j^1 \psi$ to the 1-jet prolongation of $\psi$, which locally takes the form $(x^\mu, \psi(x), k_\mu(x))$, $\mathbf{k}$ denotes the vector field that is metric-dual to $k$, and $i_g: T^*(M) \to T(M)$ to the isomorphism of cotangent spaces and tangent spaces that is define by $g$, and is locally referred to as "raising an index."

Furthermore, if one substitutes the locally correct replacement of the exactness of $p$ that is expressed by the first equation of (3.7) with the condition that it be closed, then the equations take the form:

$$dk = 0, \qquad \delta \mathbf{k} = F(j^1 \psi), \qquad \mathbf{k} = i_g k. \tag{3.8}$$

As has been discussed elsewhere by the author [**13**], system of equations of this form basically take the generalized form of an integrability condition on a field – $\psi$, in this case, – a dual integrability condition on the dual of the field – namely, $\mathbf{k}$ – and a constitutive law that relates the original field to its dual. Hence, one can think of the field $k$ as the kinematical variable and the field $\mathbf{k}$ as the dynamical variable in the dynamics of the wave. Although the constitutive law is purely geometrical in this case, nevertheless, these equations can be generalized to define the fundamental field equations for wave motion at a "pre-metric" level, in such a way that the form (3.8) follows from the dispersion law for the wave medium. Indeed, the notion of integrability seems to play an even more fundamental role in physics than perhaps the least action principle itself. (For more discussion of this see [**13**] and references cited therein.)

### 3.2 Generalized spherical coordinates on field spaces

The wave equations that we are concerned with for the remainder of this study all take the form of the Klein-Gordon equation, with the difference between them being in the vector space *V* in which the wave function takes its values, which we then call the *field space* of the wave functions.

Interestingly, the vector spaces that seem to be most interesting in the eyes of elementary wave mechanics are $V = \mathbb{C}$, $\mathbb{C}^2$, $\mathbb{C}^4$, although $\mathbb{C}^3$ certainly plays a role in quantum chromodynamics, since it supports the fundamental representation of *SU*(3). Furthermore, the groups of interest to physics that act on these spaces are *U*(1), *SU*(2), and *SU*(2; $\mathbb{C}$), which correspond to phase rotations, Euclidian rotations or isospin rotations, and Lorentz transformations, respectively.

In fact, since all of the aforementioned groups can be embedded as manifolds in the corresponding spaces, one can say that $V - \{0\}$ in each case can be represented as the manifolds $\mathbb{R}^+ \times U(1)$, $\mathbb{R}^+ \times SU(2)$, and $\mathbb{C}^* \times SL(2; \mathbb{C})$, which underlie the Lie groups $\mathbb{C}^*$, the Weyl group of *SO*(3), and *GL*(2; $\mathbb{C}$), resp. Since these spaces are also diffeomorphic to $\mathbb{R}^+ \times S^1$, $\mathbb{R}^+ \times S^3$, and $\mathbb{C}^* \times S^3_{\mathbb{C}}$, respectively, we can say that we introducing generalized spherical coordinates into $V - \{0\}$, in effect. Actually, for the sake of differential equations, we will generally be working with the corresponding coordinates on their Lie algebras $\mathbb{R} \oplus \mathfrak{u}(1)$, $\mathbb{R} \oplus \mathfrak{su}(2)$, and $\mathbb{C}^* \oplus \mathfrak{sl}(2; \mathbb{C})$. Moreover, we will find that we are really more concerned with the structure constants of the associative algebras over these vector spaces than the Lie algebras themselves.

In the case of $V = \mathbb{C}$, the spherical coordinates in question on $\mathbb{C}^* = \mathbb{R}^2 - \{0\}$ are $(R, e^{i\theta})$. The coordinates on the Lie algebra $\mathbb{R} \oplus \mathfrak{u}(1)$ are then the usual polar coordinates $(r, \theta)$ for $\mathbb{R}^2 - \{0\}$, as long as one sets $R = e^r$.

In the case of $V = \mathbb{C}^2$, since *SU*(2) is diffeomorphic to the real three-sphere $S^3$, we are essentially extending polar coordinates from $\mathbb{R}^2 - \{0\}$ to $\mathbb{R}^4 - \{0\}$, by using $(R, \exp(\phi^1 e_1), \exp(\phi^2 e_2), \exp(\phi^3 e_3))$ as our spherical coordinates. The corresponding coordinates on $\mathbb{R} \oplus \mathfrak{su}(2)$ are $(r, \phi^1, \phi^2, \phi^3)$. Since the $\phi^i$ generate the Euler angles as coordinates on *SU*(2), they are to be distinguished from the usual right ascension and codeclination angles that one uses for the spherical coordinates of $\mathbb{R}^3 - \{0\}$.

In the case of $V = \mathbb{C}^4$, since $SL(2; \mathbb{C})$ is diffeomorphic to the complex 3-sphere, we are essentially defining a "complex spherical coordinate" representation of $\mathbb{C}^4 - \{0\}$ as $\mathbb{C}^* \times S_{\mathbb{C}}^3$. since the image of $\mathfrak{sl}(2; \mathbb{C})$ under exp is $SL(2; \mathbb{C})$, which is diffeomorphic to the complex 3-sphere. The generalized spherical coordinates on $\mathbb{C}^* \times SL(2; \mathbb{C})$ are then $(R, \exp(\phi^1 e_1), \exp(\phi^2 e_2), \exp(\phi^3 e_3))$, as in the case of $\mathbb{R}^+ \times SU(2)$, except that now they are complex numbers, not real numbers. Similarly, the coordinates on the Lie algebra $\mathfrak{gl}(2; \mathbb{C}) = \mathbb{C} \oplus \mathfrak{sl}(2; \mathbb{C})$ are still of the form $(r, \phi^1, \phi^2, \phi^3)$, except that they are also complex now.

### 3.3 Wave equations in generalized spherical coordinates

We now come to the main subject of the present article, which is what happens to the form of the wave equation, in any of the above forms, when one gives the vector space $V$ of values taken by the wave functions the generalized spherical coordinates that we just introduced. We assume the $V$ is a defined over the field $\mathbb{K}$ of scalars, which we assume to be either $\mathbb{R}$ or $\mathbb{C}$, for our purposes. Note that the process of introducing generalized spherical coordinates into $V - \{0\}$ is generally distinct from the process of giving the manifold $M$ itself a system of spherical coordinates, unless, for instance, the vector space $V$ is represented by the tangent spaces to $M$. Hence, we are more concerned with symmetries of the field space than with symmetries of the spacetime manifold itself.

This process begins with assuming that the motion of the wave $\psi$ is due to the action of a Lie group $G$ on $V$, which we shall assume to be a linear action; i.e., we have a representation of $G$ as a subgroup of $GL(V)$. This has the advantage of allowing us to denote the action of a group element $A \in G$ on a vector $\mathbf{v} \in V$ by means of the multiplication of matrices for both when one chooses a basis $\{\mathbf{e}_i, i = 0, \ldots, n)$ for $V$, which is not the same thing as choosing a coordinate system for $M$. The vector $A\mathbf{v}$ will then have the components $A^i_j v^j$ relative to this basis.

Really, it is not the group $G$ that plays the key role here, but the group of all smooth functions $A: M \to G$, which then acts on the smooth functions $\psi: M \to V$ pointwise; that is:

$$(A\psi)(x) = A(x)\psi(x). \tag{3.9}$$

Furthermore, we shall be concerned only with the identity component of $G$, so that any $A \in G$ can be represented in the form $\exp(\phi^a e_a)$, where the $\phi^a$ are smooth functions with values in $\mathbb{K}$ and $\{e_a, a = 0, \ldots, n-1\}$ defines a basis for the Lie algebra $\mathfrak{g}$ of $G$. In all cases, we will be assuming that the subgroup generated by all $\exp((\phi^0 e_0)$ is the center of $G$, so the Lie algebra generated by $e_0$ is also the center of $\mathfrak{g}$. Since we are assuming that $G$ is represented by a subgroup of $GL(n; \mathbb{K})$, the Lie algebra $\mathfrak{g}$ will be represented by a subalgebra of the Lie algebra $\mathfrak{gl}(n; \mathbb{K})$. However, as we shall see, it is not only the *Lie* algebra $\mathfrak{g}$ that is generated by these basis elements that we need to deal with, but the *associative* algebra $\mathfrak{A}$, as well, which will also be a sub-algebra of $M(n; \mathbb{K})$, which is defined over the same vector space of matrices as $\mathfrak{gl}(n; \mathbb{K})$.

The starting point for our conversion of the wave equations is given by a generalization of the Heisenberg representation for wave mechanics, namely, we let by $\psi_0$ be a constant field on $M$ and define the more general field $\psi$ as:

$$\psi = \exp(\phi^a e_a)\psi_0. \qquad (3.10)$$

In this expression, the $\phi^a$ are smooth functions on $M$ with values in $\mathfrak{g}$.

The effect of this definition is to say that when one chooses a reference "phase" $\psi_0$ for the wave field $\psi$ the motion of the wave is representable in terms of the phase group elements alone. The mathematical focus then shifts from $V$ to $\mathfrak{g}$. However, the details of the particular choice of representation of $\mathfrak{g}$ will re-emerge in the consideration of bilinear covariants.

One can just as well say that we are replacing $\psi$ with $\exp(\phi^a e_a)$, since, as we will notice in what follows, the choice of $\psi_0$ plays no particular role. This amounts to saying that we are using the $\phi^a$ as generalized spherical coordinates for $V - \{0\}$. Hence, this representation of the values of the wave function $\psi$ is valid only for the support of $\psi$.

Now, apply the d'Alembertian operator to both sides of (3.10):

$$\Box \psi = \Box (\exp(\phi^a e_a))\psi_0. \qquad (3.11)$$

By straightforward computation, one finds that:

$$\Box (\exp(\phi^a e_a)) = [g^{\mu\nu} \phi^a{}_{,\mu} \phi^b{}_{,\nu} e_a e_b + g^{\mu\nu} \phi^b{}_{,\mu,\nu} e_b] \exp(\phi^a e_a). \qquad (3.12)$$

It is in the left-hand expression that we find the justification for the mathematical abstractions and generalizations that are discussed in the Appendix. One sees that is it the product $e_a e_b$ that enters into the expression, not the Lie bracket. Hence, if one knows the structure constants $a^c_{ab}$ of the associative algebra $\mathfrak{A}$ that is generated by the $e_a$ then one can re-express the term in brackets as:

$$[g^{\mu\nu} \phi^a{}_{,\mu} \phi^b{}_{,\nu} e_a e_b + g^{\mu\nu} \phi^b{}_{,\mu,\nu} e_b] = [g^{\mu\nu} a^c_{ab} \phi^a{}_{,\mu} \phi^b{}_{,\nu} + \Box \phi^c] e_c. \qquad (3.13)$$

From (3.11), we can then say that:

$$\Box \psi = \{ [\Box \phi^c + g^{\mu\nu} a^c_{ab} \phi^a{}_{,\mu} \phi^b{}_{,\nu}] e_c \} \psi, \qquad (3.14)$$

so the eigenvalue equation then gives rise to the following set of $n$ quasi-linear second-order partial differential equations for the $\phi^a$:

$$\Box \phi^c + g^{\mu\nu} a^c_{ab} \phi^a_{,\mu} \phi^b_{,\nu} = \lambda \delta^c_0. \tag{3.15}$$

Note that the nonlinear contribution to what started out as a linear equation is solely due to the nature of the algebra $\mathfrak{A}$.

We can further convert this system of $n$ second-order equations into a system of $2n$ first-order equations in the usual way:

$$\phi^a_{,\mu} = k^a_\mu, \qquad g^{\mu\nu}(k^c_{\mu,\nu} + a^c_{ab} k^a_\mu k^b_\nu) = \lambda \delta^c_0, \tag{3.16}$$

or, to put this into the quasi-linear first order form of (3.4):

$$\phi^a_{,\mu} = k^a_\mu, \qquad k^{\mu c}_{,\mu} = \lambda \delta^c_0 - g^{\mu\nu} a^c_{ab} k^a_\mu k^b_\nu, \tag{3.17}$$

By further defining the 1-forms $k^a = k^a_\mu dx^\mu$ and assuming that the $g^{\mu\nu}$ are constant for the chosen coordinate system, we can put this into coordinate-independent form (but not independent of the set of generators on $\mathfrak{A}$) as in (3.7):

$$d\phi^a = k^a, \qquad \delta \mathbf{k}^a = \lambda \delta^c_0 - a^c_{ab} g(k^a, k^b), \qquad \mathbf{k}^a = i_g k^a. \tag{3.18}$$

One can also generalize the first set of equations to put this into the form (3.8):

$$dk^a = 0, \qquad \delta \mathbf{k}^c = \lambda \delta^c_0 - a^c_{ab} g(k^a, k^b), \qquad \mathbf{k}^a = i_g k^a. \tag{3.19}$$

One must notice that these equations do not refer to the $\phi^a$ explicitly anywhere, and can thus be regarded as equations in the 1-forms $k^a$.

Since we shall be exclusively concerned with algebras that decompose into the direct sum of a 1-dimensional center and the Lie algebra of a Lie group, we shall then re-express (3.19) in terms of $k^0$ and $k^i$, explicitly:

$$dk^0 = 0, \qquad \delta \mathbf{k}^0 = \lambda - a^0_{ab} g(k^a, k^b), \qquad \mathbf{k}^0 = i_g k^0. \tag{3.20a}$$
$$dk^i = 0, \qquad \delta \mathbf{k}^i = - a^i_{ab} g(k^a, k^b), \qquad \mathbf{k}^i = i_g k^i. \tag{3.20b}$$

In the next section, we shall examine the form that these equations take for the various algebras of interest to quantum wave mechanics.

## 4 Physical examples

In order to apply the method of generalized spherical coordinates on the field space to the specific cases of $V = \mathbb{C}, \mathbb{C}^2, \mathbb{C}^4$ that are of interest to us, one need only go back to section 2 to obtain the appropriate structure constants for the algebras $\mathfrak{A} = \mathbb{C}, \mathbb{H}$, and $\mathbb{H}^{\mathbb{C}}$, respectively.

### 4.1 Klein-Gordon equation

When one chooses $\mathfrak{A} = \mathbb{C}$, the resulting equation is the usual Klein-Gordon equation:

$$\Box \psi = - k_0^2 \psi, \qquad (4.1)$$

which basically represents the eigenvalue equation for the d'Alembertian operator, under the assumption that the one is looking for negative eigenvalues[3]. In wave mechanics, where $\psi$ is a complex-valued $C^2$ function on spacetime that represents the wavefunction of a moving pointlike particle, one assumes that $k_0 = m_0 c / \hbar$ is the Compton wave number that is associated with the rest mass $m_0$ of the particle.

The extension of the Madelung transformation that was carried by Takabayasi [**2**] began with expressing $\psi$ in polar form:

$$\psi = R \, e^{iS/\hbar}, \qquad (4.2)$$

in which $R$ and $S$ are then real-valued $C^2$ functions on spacetime, and then obtaining real differential equations by annulling the real and imaginary parts of the equation that followed from (4.1) by the substitution (4.2).

We shall apply the slightly generalized results that we obtained above by the Ansatz:

$$\psi = \exp(\ln \sqrt{\rho} + i\theta)\psi_0. \qquad (4.3)$$

If we regard $\mathbb{C}$ as $\mathbb{R}^2$ then we can express this equation in the real form:

$$\psi = \exp(\ln \sqrt{\rho}\, I + \theta J)\psi_0. \qquad (4.4)$$

That is, $\phi^0 = \ln \sqrt{\rho}$, $\phi^1 = \theta$ in this case.

---

[3] The negative sign appears by starting with the d'Alembertian operator and replacing the operators $\partial/\partial x^\mu$ with their Fourier transforms $ik_\mu$ – i.e., their symbols. This associates the eigenvalue equation for the d'Alembertian with a dispersion law for the frequency-wave number 1-form $k = k_\mu dx^\mu$, namely $k^2 = k_0^2$.

We recall that the non-zero structure constants of the algebra $\mathbb{C}$ are $a_{00}^0 = - a_{11}^0 = a_{01}^1 = a_{10}^1 = 1$. Substitution in the general equations (3.20a, b) gives the pair of equations:

$$dk^0 = 0, \qquad \delta\mathbf{k}^0 = -k_0^2 - g(k^0, k^0) + g(k^1, k^1), \qquad \mathbf{k}^0 = i_g k^0. \qquad (4.5a)$$
$$dk^1 = 0, \qquad \delta\mathbf{k}^1 = -2g(k^0, k^1), \qquad \mathbf{k}^1 = i_g k^1. \qquad (4.5b)$$

which generalize the real and imaginary equations.

If we set:

$$k^0 = d(\ln\sqrt{\rho}) = \frac{d\sqrt{\rho}}{\sqrt{\rho}} = \frac{d\rho}{2\rho}, \qquad k^1 = d\theta \equiv k \qquad (4.6)$$

and take into account that if $f$ is a 0-form and and $\alpha$ is a 1-form then one has:

$$\delta(f\alpha) = g(df, \alpha) + f\delta\alpha, \qquad (4.7)$$

then we can further put (4.5a, b) into the form:

$$dk = 0, \qquad \delta\mathbf{J} = 0, \qquad \mathbf{J} \equiv i_{\rho g} k, \qquad k^2 = k_0^2 + \frac{\Box\sqrt{\rho}}{\sqrt{\rho}}. \qquad (4.8)$$

This is essentially an extension of our canonical set of equations (3.8) to include a dispersion law. Moreover, we are no longer using the metric $g$ alone to transform cotangent vectors into tangent vectors, but a metric conformal to it by way of the conformal factor $\rho$.

In the Madelung-Takabayasi "hydrodynamical" interpretation of these resulting equations, $\rho = \psi^*\psi$, is assumed to be a particle number density function for an extended matter distribution. (Although the term "fluid" is usually used here, since the resulting stress tensor is not actually that of an isotropic medium, and the anisotropy does not seem to be due to viscosity, we shall avoid that convention.) If one lets $\theta = S/\hbar$ then:

$$k = d\theta = dS/\hbar = (1/\hbar)p, \qquad (4.9)$$

in which $p$ represents the energy-momentum density 1-form of the motion then (4.8) become:

$$dp = 0, \qquad \delta\mathbf{J} = 0, \qquad \mathbf{J} \equiv i_{\rho g} p = \rho\mathbf{p}, \qquad p^2 = \left(m_0^2 + \hbar^2 \frac{\Box\sqrt{\rho}}{\sqrt{\rho}}\right)c^2. \qquad (4.10)$$

The usual argument for regarding the second equation as the law of conservation of mass is based on the assumption that $p/m_0$ can be given the interpretation of the covelocity 1-form $u$ of the motion. However, this argument is conceptually flawed by the fact that $m_0$ is the *total* rest mass of the distribution, and a more natural relation between $p$ and $u$ is defined by means of the rest mass *density* $\rho_0 \equiv m_0\rho$, namely:

$$p = \rho_0 u. \tag{4.11}$$

However, the overall effect on equations (4.10) is moot if one sets $\mathbf{p} = m_0\mathbf{u}$, since that makes $\mathbf{J} = \rho_0\mathbf{u}$, which then represents the energy-momentum vector field.

This puts the middle equation into the form:

$$\delta(\rho_0\mathbf{u}) = 0, \tag{4.12}$$

which then says either that the motion is dynamically incompressible[4] or, equivalently, that rest mass is conserved along the flow of $\mathbf{u}$.

With the assignment (4.11), the first equation in (4.11) does not say that the flow is kinematically irrotational, which amounts to the vanishing of $du$, but that its dynamic vorticity $dp$ vanishes. Its kinematic vorticity $du$ obeys:

$$du = - d(\ln \rho_0) \wedge u . \tag{4.13}$$

That is, the flow is kinematically irrotational if and only if the gradient $d\rho_0$ of the rest mass density is collinear with the covelocity.

As for the last equation in (4.10), it usually gets the interpretation of the Hamilton-Jacobi equation that is associated with the Hamiltonian:

$$H = T + U_\hbar = p^2 c^2 - \hbar^2 c^2 \frac{\Box\sqrt{\rho}}{\sqrt{\rho}}, \tag{4.14}$$

as long as one thinks of the somewhat enigmatic term:

$$U_\hbar = -\hbar^2 c^2 \frac{\Box\sqrt{\rho}}{\sqrt{\rho}} \tag{4.15}$$

as a "quantum potential," presumably because the factor $\hbar^2$ gives it the character of a small correction term to the total energy that vanishes in the classical limit. Of course, this then begs the question of where it comes from independently of the Madelung transformation.

One should once observe that since $m_0$ is the total rest mass, it is also probably incorrect to assume that $p^2 = m_0^2 c^2$, which makes perfect sense for pointlike matter rather than assuming that:

$$p^2 = \rho_0^2 c^2 , \tag{4.16}$$

in which we are dealing with the rest mass density $\rho_0$, instead.

---

[4] The terminology of "kinematic" and "dynamic" quantities used here is that of Carter [**14**]. Kinematic quantities pertain to the covelocity 1-form $u$, while dynamic quantities pertain to the energy-momentum 1-form $p$.

From (4.10), we should have:

$$\rho_0^2 = m_0^2 + \hbar^2 \frac{\Box \sqrt{\rho}}{\sqrt{\rho}}. \tag{4.17}$$

When the rest frequency $k_0$ is no longer a constant, but a function on spacetime, one sees that the Klein-Gordon equation itself takes on a subtly more complicated form:

$$\Box \psi = -k_0^2(x^\mu)\psi \tag{4.18}$$

Hence, the most natural wave equation for extended matter distributions is more general than the usual eigenvalue equation for the d'Alembertian, since the eigenvalues are *functions*, rather than constants. The appearance of constants would have to follow from the process of reducing from an extended body to a pointlike body; e.g., from mass density to total mass.

One can also take the position that the dispersion law for $k$ is what is most fundamental and that a more general form that includes the higher moments of $\rho$ involves the generalization of $k_0$ to some function $k_0(k)$. However, we shall defer a detailed analysis of this possibility for a subsequent study, since it also implies that the Klein-Gordon equation might require a different modification than the one suggested in (4.18).

### 4.2 Relativistic Pauli equation

Actually, what one refers to as the "Pauli equation" is the non-relativistic wave equation that one obtains from the Schrödinger equation by substituting wave functions that take their values in $\mathbb{C}^2$, which carries a representation of $SU(2)$, as a first step towards including spin in the wave functions, such as one would desire for electrons, for instance.

However, since the relativistic form of the Schrödinger equation, namely, the Klein-Gordon equation, is mathematically more concise to deal with, although it is physically more debatable in its interpretation, we shall proceed from the Klein-Gordon equation (4.1) with the wave function $\psi$ taking its values in the algebra $\mathfrak{A} = \mathbb{H}$.

Hence, we make the Ansatz:

$$\psi = \exp(\ln \sqrt{\rho}\ e_0 + \phi^i e_i)\psi_0 = \sqrt{\rho}\ \exp(\phi^i e_i)\psi_0. \tag{4.19}$$

Thus we use $\phi^0 = \ln \sqrt{\rho}$, as before, along with the three $\phi^i$ as the components of the wave function $\psi$ for this choice of $\mathfrak{A}$.

The real form that the Pauli equation takes as a result of this substitution is given by (3.20a,b). If we recall that the structure constants of $\mathbb{H}$ are given by (2.18) $a_{0\beta}^\gamma = 1/2\, b_{0\beta}^\gamma = \delta_\beta^\gamma$, $a_{ij}^0 = 1/2\, b_{ij}^0 = -\delta_{ij}$, $a_{ij}^k = 1/2\, c_{ij}^k = \varepsilon_{ijk}$, in which we now index the basis elements for $\mathbb{H}$ by $\alpha, \beta, \gamma = 0, \ldots, 3$ to distinguish them from the coordinates of $M$, then the (3.20a,b) give the following set of four real second order quasi-linear partial differential equations for the $k^\alpha = d\phi^\alpha$:

$$dk^0 = 0, \quad \delta\mathbf{k}^0 = -k_0^2 - \eta_{\alpha\beta}\, g(k^\alpha, k^\beta), \quad \mathbf{k}^0 = i_g k^0. \tag{4.20a}$$
$$dk^i = 0, \quad \delta\mathbf{k}^i = -2g(k^0, k^i), \quad \mathbf{k}^i = i_g k^i. \tag{4.20b}$$

In the middle equation of (4.20b), we have taken into account that the symmetry of $g$ clashes with the anti-symmetry of $\varepsilon$ to annul the corresponding term of the form $\varepsilon_{ijk}\, g(k^j, k^k)$.

If we make the same substitution $k^0 = d(\ln\sqrt{\rho})$ as we did for the case of the algebra $\mathbb{C}$ then these equations take the form:

$$dk^i = 0, \quad \delta\mathbf{J}^i = 0. \quad \mathbf{J}^i = i_{\rho g} k^i \quad \delta_{ij}\, g(k^i, k^j) = k_0^2 + \frac{\Box\sqrt{\rho}}{\sqrt{\rho}}, \tag{4.21}$$

Now, we see that we have three currents $\mathbf{J}^i$ to deal with, not just the one, and rather than a separate dispersion law for each $k^i$, we have a single equation for all three.

However, one must realize that collectively the three 1-forms $k^i$ define the components of a 1-form $K$ on $M$ with values in $\mathbb{R}^3$, relative to its canonical basis $\boldsymbol{\delta}^i$, namely:

$$K = k^i \otimes \boldsymbol{\delta}_i. \tag{4.22}$$

Thus, one can find another orthonormal basis $\mathbf{e}_i(x)$ for $\mathbb{R}^3$ that generally depends upon which point $x \in M$ one considers and differs from $\boldsymbol{\delta}_i$ by a rotation $\tilde{R}^i_j(x)$, so:

$$\mathbf{e}_j(x) = \tilde{R}^i_j(x)\, \boldsymbol{\delta}_i, \tag{4.23}$$

and has the property that it is adapted to $K$ in that:

$$K = k \otimes \mathbf{e}_1 = R^1_i k^i \otimes \mathbf{e}_1. \tag{4.24}$$

in which $k$ is a single 1-form on $M$.

When one takes the exterior derivative of $k$, one obtains, taking into account that each $k^i$ is closed:

$$dk = dR^1_i \wedge k^i = \omega^1_1 \wedge k = 0, \tag{4.25}$$

in which we have defined the angular velocity of the frame $\mathbf{e}_i$ relative to $\boldsymbol{\delta}_i$ by:

$$\omega^i_j = dR^i_k\, \tilde{R}^k_j, \tag{4.26}$$

which is then anti-symmetric, since it is an element of $\mathfrak{so}(3; \mathbb{R})$.

If we then define:

$$\mathbf{J} = i_{\rho g} k = R_i^1 (i_{\rho g} k^i) = R_i^1 \mathbf{J}^i, \tag{4.27}$$

we see, taking into account that $\delta \mathbf{J}^i$ vanishes for each $i$, that:

$$\delta \mathbf{J} = g(dR_i^1, \mathbf{J}^i) = g(\omega_1^1, \mathbf{J}) = 0. \tag{4.28}$$

Furthermore, we can see that:

$$\delta_{ij} g(\tilde{R}_1^i k, \tilde{R}_1^j k) = (\delta_{ij} \tilde{R}_1^i \tilde{R}_1^j) g(k,k) = k^2, \tag{4.29}$$

which also follows from the fact that $R$ is orthogonal.

Summarizing the results of (4.25), (4.27), (4.28), and (4.29), we see that we can put (4.21) into the equivalent form:

$$dk = 0, \qquad \delta \mathbf{J} = 0, \qquad \mathbf{J} = i_{\rho g} k, \qquad k^2 = k_0^2 + \frac{\Box \sqrt{\rho}}{\sqrt{\rho}}, \tag{4.30}$$

which is precisely the same as (4.8).

However, one aspect of this model that demands further interpretation is the fact that although we started out with an action of $SU(2)$ on the field space, nevertheless, the structure constants of its Lie algebra disappeared from the real form of the field equations identically. Similarly, in a "co-spinning" frame on the field space, the fundamental equations have not changed by the expansion of the algebra of the field space. The question then arises of what happened to this "spin" of the resulting wave. We shall defer speculation on this problem of interpretation, along with that of the quantum potential, until after we have treated the role of spin in the field space in a subsequent study.

### 4.3 Bi-Dirac equation

What we are calling the *bi-Dirac* equation here is simply the Klein-Gordon equation when the wave function $\psi$ takes its values in $\mathbb{C}^4$. The Dirac equation, properly speaking, is obtained when one essentially takes the "square root" of the operators that act on $\psi$ in order to reduce the order of the differential equations from two to one:

$$\slashed{\nabla} \psi \equiv \gamma^\mu \partial_\mu \psi = i k_0 \psi. \tag{4.31}$$

As usual, the reduction of order is associated with an expansion in the number of equations, which must now include the complex conjugate equations:

$$\slashed{\nabla} \psi^* = -i k_0 \psi^*. \tag{4.32}$$

As $\square = \overline{\nabla}^2$, one sees that in order for $\psi$ to be a solution of the Dirac equation, it is necessary that it also be a solution of the Klein-Gordon equation, but not sufficient. Since solutions of the (zero-eigenvalue) equation that is defined by the square of the Laplacian operator are called "biharmonic," we shall refer to solutions of the Klein-Gordon equation that take their values in $\mathbb{C}^4$ as "bi-Dirac spinor wavefunctions."

For our present purposes, the advantage of using the Dirac equation in place of the Klein-Gordon – viz., a reduction of the order of the equations – is not enough to outweigh the confusion created by introducing a representation of an algebra that we are not immediately concerned with when it also overlaps the vector space of the one that are dealing with. Hence, we shall content ourselves with examining the consequence of representing the space $\mathbb{C}^* - \{0\}$ in complex spherical form by means of $\mathbb{C}^* \times SL(2; \mathbb{C})$ and seeing what happens to our generalized wave equations (3.15) when we use the structure constants of the algebra $\mathbb{H}^\mathbb{C}$.

Since $\mathbb{H}^\mathbb{C}$ is the complexification of $\mathbb{H}$, if one allows the functions $\phi^\alpha$, $\alpha = 0, \ldots, 3$, along with the 1-forms $k^\alpha$, to take their values in $\mathbb{C}$ then nothing changes in the general form of the equations (3.20a,b) that we derived for the relativistic Pauli equation. However, if one wishes for all of the wave functions to be real-valued then one must extend the set of four equations to a set of eight equations for the $k^\alpha$ and $\bar{k}^\alpha$ by including the other structure constants for $\mathbb{H}^\mathbb{C}$ when one regards it as real and eight-dimensional.

Recall that the full set of non-zero structure constants for $\mathbb{H}^\mathbb{C}$ were given by (2.18) and (2.37) to be $a^\kappa_{0\nu} = 1/2\, b^\kappa_{0\nu} = \delta^\kappa_\nu$, $a^0_{ij} = 1/2\, b^0_{ij} = -\delta_{ij}$, $a^k_{ij} = 1/2\, c^k_{ij} = \varepsilon_{ijk}$, $a^{\bar\kappa}_{\mu\bar\nu} = a^{\bar\kappa}_{\bar\mu\nu} = -a^\kappa_{\bar\mu\bar\nu} = a^\kappa_{\mu\nu}$. Substitution into (3.20a, b) gives:

$$dk^0 = 0, \qquad \delta\mathbf{k}^0 = -k_0^2 - \eta_{\alpha\beta}\, g(k^\alpha, k^\beta) + \eta_{\alpha\beta}\, g(\bar{k}^\alpha, \bar{k}^\beta), \qquad \mathbf{k}^0 = i_g k^0, \qquad (4.33a)$$

$$dk^i = 0, \qquad \delta\mathbf{k}^i = -2g(k^0, k^i) + 2g(\bar{k}^0, \bar{k}^i), \qquad \mathbf{k}^i = i_g k^i, \qquad (4.33b)$$

$$d\bar{k}^0 = 0, \qquad \delta\bar{\mathbf{k}}^0 = -2\eta_{\alpha\beta}\, g(k^\alpha, \bar{k}^\beta), \qquad \bar{\mathbf{k}}^0 = i_g \bar{k}^0, \qquad (4.33c)$$

$$d\bar{k}^i = 0, \qquad \delta\bar{\mathbf{k}}^i = -2g(k^0, \bar{k}^i) - 2g(\bar{k}^0, k^i), \qquad \bar{\mathbf{k}}^i = i_g \bar{k}^i. \qquad (4.33d)$$

Note that one could also have obtained these from (4.20a, b) by substituting $k^\alpha + i\bar{k}^\alpha$ for $k^\alpha$ and separating the real and imaginary equations. That is, although we are representing $\mathbb{C}^* - \{0\}$ by complex spherical coordinates, nevertheless, the algebra $\mathbb{H}^\mathbb{C}$, which exponentiates onto it, is still represented by Cartesian coordinates.

By the substitution of $k^0 = d(\ln\sqrt{\rho})$, we can then put (4.33a-d) into the form:

$$\delta_{ij} [g(k^i, k^j) - g(\bar{k}^i, \bar{k}^j) = k_0^2 - (\bar{k}^0)^2 + \frac{\Box \sqrt{\rho}}{\sqrt{\rho}}, \tag{4.34a}$$

$$dk^i = 0, \qquad \delta \mathbf{J}^i = \frac{2}{\rho} g(\bar{\mathbf{J}}^0, \bar{\mathbf{J}}^i), \qquad \mathbf{J}^i = i_{\rho p} k^i, \tag{4.34b}$$

$$d\bar{k}^0 = 0, \qquad \delta \bar{\mathbf{J}}^0 = -\frac{2}{\rho} \delta_{ij} g(\mathbf{J}^i, \bar{\mathbf{J}}^j), \qquad \bar{\mathbf{J}}^0 = i_{\rho g} \bar{k}^0, \tag{4.34c}$$

$$d\bar{k}^i = 0, \qquad \delta \bar{\mathbf{J}}^i = -\frac{2}{\rho} g(\bar{\mathbf{J}}^0, \mathbf{J}^i), \qquad \bar{\mathbf{J}}^i = i_{\rho g} \bar{k}^i. \tag{4.34d}$$

We can do the same thing with $k^i$ and $\bar{k}^i$ that we did with $k^i$ for the relativistic Pauli equation, namely, define an adapted frame, except that now we must be careful to recognize that generally we must define *two* independent orthonormal 3-frames $\mathbf{e}_i = R_i^j \delta_j$ and $\mathbf{f}_i = \bar{R}_i^j \delta_j$ in $\mathbb{R}^3$ that are adapted to the vectors $\mathbf{k} = k^i \delta_i$ and $\bar{\mathbf{k}} = \bar{k}^i \delta_i$, resp. Note that a possible source of confusion is that although these two real orthonormal 3-frames – or rather, $\mathbf{e}_i$ and $i\mathbf{f}_i$ – belong to separate subspaces of the complex field space, we are nonetheless defining both of them in the *same* real vector space. Hence, in particular, the scalar products $\delta(\mathbf{e}_i, \mathbf{f}_j)$ do not all have to vanish, when we let $\delta = \delta_{ij} \theta^i \otimes \theta^j$ ($\theta^i(\mathbf{e}_j) = \delta^i_j$) represent the Euclidian scalar product on $\mathbb{R}^3$. In fact, since both $\mathbf{e}_i$ and $\mathbf{f}_i$ are both real orthonormal, one has:

$$\delta(\mathbf{e}_i + i\mathbf{f}_i, \mathbf{e}_j + i\mathbf{f}_j) = 2i \, \delta(\mathbf{e}_i, \mathbf{f}_i), \tag{4.35}$$

which shows that the set $\{\mathbf{e}_i, i\mathbf{f}_i\}$ of six real 3-vectors does not define a complex orthonormal frame.

We then define the following 1-forms and vector fields on $M$:

$$k = R_j^1 k^j, \qquad \bar{k} = \bar{R}_i^1 \bar{k}^i, \qquad \mathbf{J} = i_{\rho g} k = R_i^1 \mathbf{J}^i, \qquad \bar{\mathbf{J}} = i_{\rho g} \bar{k} = \bar{R}_i^1 \bar{\mathbf{J}}^i, \tag{4.36}$$

and note that:

$$R_i^1 \tilde{R}_1^i = \delta_{ij} \bar{R}_1^i R_1^j = \delta(\mathbf{e}_1, \mathbf{f}_1). \tag{4.37}$$

We can then put our basic system of real equations into the form:

$$k^2 - \bar{k}^2 = k_0^2 - (\bar{k}^0)^2 + \frac{\Box \sqrt{\rho}}{\sqrt{\rho}}, \tag{4.38a}$$

$$dk = 0, \qquad \delta \mathbf{J} = \frac{2}{\rho} \delta(\mathbf{e}_1, \mathbf{f}_1) g(\bar{\mathbf{J}}_0, \bar{\mathbf{J}}), \qquad \mathbf{J} = i_{\rho p} k, \tag{4.38b}$$

$$d\bar{k}^0 = 0, \qquad \delta\bar{\mathbf{J}}^0 = -\frac{2}{\rho}\delta(\mathbf{e}_1,\mathbf{f}_1)g(\mathbf{J},\bar{\mathbf{J}}), \qquad \bar{\mathbf{J}}^0 = i_{\rho g}\bar{k}^0, \qquad (4.38c)$$

$$d\bar{k} = 0, \qquad \delta\bar{\mathbf{J}} = -\frac{2}{\rho}\delta(\mathbf{e}_1,\mathbf{f}_1)g(\bar{\mathbf{J}}_0,\mathbf{J}), \qquad \bar{\mathbf{J}} = i_{\rho g}\bar{k}. \qquad (4.38d)$$

One must avoid the temptation to confuse the scalar product $g$, which is defined on the tangent spaces to $M$ and the scalar product $\delta$, which is defined on the field space. However, since a 3-frame in a tangent space $T_xM$ can be redefined to be the image of a 3-frame in $\mathbb{R}^3$ under a linear injection of $\mathbb{R}^3$ in $T_xM$, it is possible to relate the two scalar products, at least locally.

One sees that the quantity $\delta(\mathbf{e}_1, \mathbf{f}_1)$ plays a key role in equations (4.38a, b, c, d) since its vanishing would imply the conservation of all three of the currents $\bar{\mathbf{J}}^0$, $\mathbf{J}$, and $\bar{\mathbf{J}}$.

Although it is most commonplace for physics to represent geometrical objects, such as $\mathbf{e}_i$ and $\mathbf{f}_i$, in the tangent spaces to the spacetime manifold, we must nonetheless point out that the dispersion law (4.38a) seems to be more naturally associated with the scalar product $(F \wedge {}^*F)(\mathbf{V}) = E^2 - B^2$ that one defines on 2-forms by means of the Hodge operator $*$ than with the scalar product on Minkowski space itself. This is simply due to the fact that the Cartan-Killing scalar product on $\mathfrak{so}(3; \mathbb{C})$ is isometric to the complex Euclidian scalar product on $\mathbb{R}^3$, and when one gives the vector space $\Lambda^2(\mathbb{R}^4)$ a complex structure by way of a $*$ isomorphism that obeys $*^2 = -I$ the vector space of such 2-forms becomes $\mathbb{C}$-linear isomorphic to the vector space underlying $\mathfrak{so}(3; \mathbb{C})$. Hence, we see that it is probably physically incorrect to represent the field space in terms of the tangent spaces to $M$ directly, since the field space is more directly related to the Lie algebra $\mathfrak{so}(3,1)$ than it is to Minkowski space.

### 4.4 Dirac equation

At first glance, it would seem that the Dirac equation (4.31) would admit a natural transformation under the imposition of spherical coordinates on $\mathbb{C}^4$, since the Dirac $\gamma$ matrices belong to the matrix algebra $M(4; \mathbb{C})$, along with the $E_\mu$ and $\bar{E}_\mu$ matrices that we have been using to define the spherical coordinates.

However, on closer inspection, one notices that we are dealing with two different actions of the matrices in $M(4; \mathbb{C})$ on $\mathbb{C}^4$ that refer to two different representations of the points of $\mathbb{C}^4$. In the case of the Dirac equation, one is essentially regarding the points of $\mathbb{C}^4$ in Cartesian coordinates, whereas in the case of generalized spherical coordinates, those points are associated with matrices in $M(4; \mathbb{C})$. Although the action of matrices in $M(4; \mathbb{C})$ is defined on both column vectors in $\mathbb{C}^4$ and the other matrices of $M(4; \mathbb{C})$, nevertheless, the effect is not the same. Hence, it would incorrect to represent the values of the wave function $\psi$ in terms of matrices and then multiply them by the Dirac matrices as if they were column vectors.

The conventional way of representing the Dirac equation by an equivalent set of real equations (see, e.g., Takabayasi [**4**]) is to use the method of *bilinear covariants*. However, a thorough description of this method and the search for any possible relationship with the present method of generalized spherical coordinates is of sufficient scope as to warrant a separate treatment from the present one. Hence, we shall defer that analysis to a subsequent article.

## 5  Discussion

In this article, we have shown that the most common wave functions of elementary quantum mechanics, both non-relativistic and relativistic, take their values in complex vector spaces – namely, $\mathbb{C}$, $\mathbb{C}^2$, and $\mathbb{C}^4$ – that all share the common property that the complement of the origin can be represented as the product manifold of a line and a sphere. In the case of $\mathbb{C}$ and $\mathbb{C}^2$, the line is $\mathbb{R}^+$ and the spheres are $S^1$ and $S^3$, respectively; in the case of $\mathbb{C}^4$, the line is $\mathbb{C}^*$ and the sphere is $S^3_{\mathbb{C}}$.

In all cases, the spheres are the underlying manifolds of Lie groups, namely, $U(1)$, $SU(2)$, and $SL(2; \mathbb{C})$. Hence, one can regard the punctured complex vector spaces as diffeomorphic to the manifold of the groups $\mathbb{R}^+ \times U(1)$, $\mathbb{R}^+ \times SU(2)$, and $\mathbb{C}^* \times SL(2; \mathbb{C}) \cong GL(2; \mathbb{C})$, resp. Their Lie algebras are then associated with the algebras of $\mathbb{C}$, $\mathbb{H}$, and $\mathbb{H}^{\mathbb{C}}$, resp., by way of the commutator bracket of the algebra product. Interestingly, other research into the nature of the Madelung-transformed wave equations of wave mechanics has indicated that they are closely related to the geometry of Weyl-Cartan spaces, which deals with connections on the bundle of Weyl frames over a Riemannian or pseudo-Riemannian manifold. (See, for instance, Delphenich [**15**], Carroll [**16**], or Shojai and Shojai [**17**].)

If we let the wave function in the generalized Klein-Gordon wave function take its values in any of these aforementioned spaces then we find that the resulting system of real differential equations involves the structure constants of the associative algebra that gives the Lie algebra in question by way of the commutator bracket. We proceeded to derive the specific forms for these systems of equations when one substitutes the specific structure constants.

In the cases of $\mathbb{C}$ and $\mathbb{H}$, the resulting system of equations can be put into the form of an integrability condition on a 1-form *k*, a constitutive law that associates it with a vector field **J**, a conservation law for **J**, and a dispersion law for *k*, at least when one passes to a co-spinning frame in the field space. In the case of $\mathbb{H}^{\mathbb{C}}$, one must also introduce 1-forms $\bar{k}^0$ and $\bar{k}$, along with associated currents $\bar{\mathbf{J}}^0$ and $\bar{\mathbf{J}}$, as a result of the complexification, and the conservation laws for the three currents involve possibly non-vanishing cosines of the angles between the three currents. Furthermore, the dispersion law also seems to be more relevant to the scalar product on $\mathfrak{so}(3; \mathbb{C})$ than the one on Minkowski space.

To summarize, there are four fundamental issues that we have deferred to later efforts, due to the length of the present paper:

*i*) The role of spin in the aforementioned systems of equations and the equation that it adds to the system.

*ii*) The relationship of the equations that we obtained in the cases of $\mathbb{C}^2$ and $\mathbb{C}^4$ to those obtained from the Pauli and Dirac equations by means of the method of bilinear covariants.

*iii*) The issue of whether the most physically meaningful extension of the theory from point particles to extended ones is to replace the constant eigenvalue $k_0$ in the Klein-Gordon equation with functions of spacetime position or to replace the $k_0$ in the dispersion relation $k^2 = k_0^2$ with a function of $k$.

*iv*) The ultimate discussion of the nature of the problem of associating the general system of equations with various physical interpretations.

As far as the last issue is concerned, we regard the essential problem as that of interpreting the nature of the density function that one obtains from the wave function by way of $\psi^\dagger \psi$. The leading possibilities then seem to be a probability density function for a point particle, a particle number density function for an extended particle, and an electric charge density function for an extended particle. These three possibilities then lead to the statistical, hydrodynamic, and electromagnetic interpretations for wave mechanics.

What seems to be emerging from the generality presented above is simply the idea that since the same mathematical formalism may be applied to otherwise unrelated physical phenomena, the problem of starting with a mathematical formalism and looking for its point of application in physics is fundamentally ambivalent. One is essentially trying to find an "optimal" section of a projection, namely, all of the physical phenomena that can be described by the same mathematical formalism.

It is the author's hope that perhaps by generalizing the nature of converting the Klein-Gordon equation into a set of real equations one can gain a clearer idea of what exactly would constitute such an optimal section when one starts with the original problem of wave mechanics, namely, the structure of the electron/positron at the elementary level.

## Appendix - Associative algebras

It might prove convenient for some readers to introduce a few elementary notions from the general theory of algebras that we shall apply in the discussion above.

### A.1 Basic concepts

An *algebra* over a vector space $V$ is a bilinear map $V \times V \to V$, $(a, b) \mapsto ab$, which is then regarded as a form of "vector multiplication." This bilinearity implies that:

$$(a + b)c = ac + bc, \qquad a(b + c) = ab + ac, \qquad (\lambda a)b = a(\lambda b) = \lambda(ab), \tag{A.1}$$

for all $a$, $b$, $c \in V$, and all scalars $\lambda \in \mathbb{K}$; for our purposes, the field $\mathbb{K}$ will either be $\mathbb{R}$ or $\mathbb{C}$. Since $V$ already has the structure of an Abelian group under vector addition, the bilinearity of the algebra product can be regarded as the left and right distributivity of the multiplication over the addition. Hence, an algebra is also a ring whose underlying set consists of the vectors in a vector space.

Examples of algebras include:

*i*) The real or complex numbers, when given the multiplication of such numbers.

*ii*) The vector space $\mathbb{R}^3$, when given the vector cross product.

*iii*) The vector space $M(n; \mathbb{K})$, which consists of all $n \times n$ matrices with entries in the field of scalars $\mathbb{K}$, with the usual multiplication of matrices.

Other examples include Lie algebras, Clifford algebras, division algebras, and Cayley algebras. Lie algebras include the second example above, which is isomorphic to the Lie algebra $\mathfrak{so}(3; \mathbb{R})$ of infinitesimal generators of three-dimensional Euclidian rotations.

The multiplication is not assumed to be associative or commutative, in general, nor is there necessarily a unity element, and certainly not every element of $V$ is assumed to have a multiplicative inverse if there is a unity element. The examples of Lie algebras and Cayley algebras are both non-associative and non-Abelian, in general.

One must clearly distinguish be a set of *basis elements* for the vector space $V$ that an algebra $\mathfrak{A}$ is defined over and a set of *generators* for the algebra $\mathfrak{A}$. Specifically, whereas a basis $\{\mathbf{e}_i, i\ 1, \ldots, n\}$ for the vector space has property that every vector in $V$ can be expressed a linear combination of the basis elements, a set of generators $\{e_a, a = 1, \ldots, m\}$ for the algebra $\mathfrak{A}$ has the property that every element of $\mathfrak{A}$ can be expressed as linear combination of finite powers of the generators. This implies that $m \leq n$, in general. For instance, since $i^2 = -1$, although 1 and $i$ are linearly independent when $\mathbb{C}$ is regarded as a two-dimensional real vector space, so $\{1, i\}$ represents a basis for the vector space, nonetheless, it is clear that $\{i\}$ is a set of generators for the algebra.

When the number of generators of an algebra is less than the dimension of the vector space that is defined over, one can usually define a basis for the vector space in terms of finite products of the generators. (In general, one can only use linear combinations of them.) For instance, the basis $\{1, i\}$ for $\mathbb{C}$ can be expressed in the form $\{i, -i^2\}$.

### A.2 Polarization

In general, one can polarize the product of any two elements $a$ and $b$ of an algebra over $V$ into a commutative part and an anti-commutative part:

$$ab = \tfrac{1}{2}(ab + ba) + \tfrac{1}{2}(ab - ba) \equiv \tfrac{1}{2}\{a, b\} + \tfrac{1}{2}[a, b], \tag{A.2}$$

in which the expression {*a*, *b*} is called the *anti-commutator* of *a* and *b* and the expression [*a*, *b*] is called the *commutator* of *a* and *b*, or also the *Lie bracket* of *a* and *b*. If *a* and *b* commute then [*a*, *b*] vanishes, whereas, if they anti-commute then {*a*, *b*} vanishes.

The anti-commutator bracket defines a commutative algebra over *V* that we shall call the *symmetric algebra* associated with the associative algebra over *V* that we started with. If that associative algebra also has a unity element then if one includes the factor of ½ in the anti-commutator bracket the resulting algebra is commutative and has a unity element.

The case of a *Clifford algebra* over an orthogonal space (*V*, *g*) is defined by imposing the following condition on the anti-commutator of elements in *V*:

$$\{a, b\} = 2g(a, b)\, e, \tag{A.3}$$

where *e* is the unity element of the algebra.

Clearly, the commutator bracket allows one to associate a Lie algebra with any associative algebra, although the Lie algebra itself is not generally associative, since the Jacobi identity measures the non-associativity of a Lie algebra. The examples of the linear Lie algebras, whose elements are the infinitesimal generators of one-parameter families of invertible linear transformations of the vector spaces $\mathbb{R}^n$ or $\mathbb{C}^n$, are all obtained by starting with sub-algebras of the associative algebra with unity $M(n, \mathbb{K})$ and giving them the Lie bracket.

Note that one can always associate a Lie algebra with any Clifford algebra by way of the commutator [*a*, *b*], but the restriction (A.3) that was placed on the product *ab* does not actually imply that this Lie algebra will be the orthogonal Lie algebra of (*V*, *g*); i.e., the one associated with infinitesimal generators of one-parameter families of orthogonal transformations.

### A.3 Structure constants

When *V* is *n*-dimensional and given a basis {$\mathbf{e}_i$, $i = 1, \ldots, n$}, from the bilinearity of the product for any algebra $\mathfrak{A}$ over *V*, it is sufficient to know all of the expressions $\mathbf{e}_i \mathbf{e}_j$ in order to determine the products of more general elements. Since these elements will all be in *V* they can be expressed in terms of the chosen basis in the form:

$$\mathbf{e}_i \mathbf{e}_j = a_{ij}^k \mathbf{e}_k \tag{A.4}$$

and the scalar constants $a_{ij}^k$ are called the *structure constants* of the algebra.

By polarizing $\mathbf{e}_i \mathbf{e}_j$, we get:

$$\mathbf{e}_i \mathbf{e}_j = \tfrac{1}{2}\{\mathbf{e}_i, \mathbf{e}_j\} + \tfrac{1}{2}[\mathbf{e}_i, \mathbf{e}_j] \equiv \tfrac{1}{2} b_{ij}^k \mathbf{e}_k + \tfrac{1}{2} c_{ij}^k \mathbf{e}_k = \tfrac{1}{2}(b_{ij}^k + c_{ij}^k)\mathbf{e}_k, \tag{A.5}$$

in which the structure constants $b_{ij}^k$ and $c_{ij}^k$ belong to the symmetric algebra and the Lie algebra associated with the given algebra, respectively. This gives us the relation:

$$a_{ij}^k = \tfrac{1}{2}(b_{ij}^k + c_{ij}^k). \tag{A.6}$$

Given such a basis for an algebra $\mathfrak{A}$, one can then express the components of the product $ab$ of two elements $a, b \in \mathfrak{A}$ in terms of the effect of the structure constants on their components with respect to the basis:

$$(ab)^k = a_{ij}^k a^i b^j. \tag{A.7}$$

As a special case of this one can see that:

$$(a^2)^k = a_{ij}^k a^i a^j = b_{ij}^k a^i a^j, \tag{A.8}$$

since the square of any element clearly represents a symmetric product and one sums over the indices of $a$ twice.

### A.4 Polynomials

Associative algebras are especially important, since they allow one to make sense of powers of algebra elements. For instance:

$$a^3 = a(aa) = (aa)a \tag{A.9}$$

makes sense only when one has the associativity of the multiplication.

One sees that by iterating (A.8) one can express any power of $a$ in terms of a homogeneous polynomial in the components $a^i$ and the symmetric structure constants $b_{ij}^k$. For instance:

$$(a^3)^k = [a(a^2)]^k = b_{il}^k b_{jm}^l a^i a^j a^m. \tag{A.10}$$

Higher powers will take an analogous form.

One can define the notions of finite and formal power series in the elements of an associative algebra because polynomial expressions of the form:

$$P(a) = \sum_{n=0}^{N} \alpha_n a^n, \qquad \alpha \in \mathbb{K},\, a \in V, \tag{A.11}$$

which will make rigorous mathematical sense for an associative algebra. (Here, if we are also assuming the existence of a unity element $I$ then we define $a^0 = I$.) Polynomials in more than one independent variable are likewise defined, although one must respect the order of multiplication in the individual terms.

If the associative algebra $\mathfrak{A}$ is defined over a finite-dimensional vector space $V$ then $\mathfrak{A}$ has finite set of basis elements $\{e_i, i = 1, \ldots, n\}$ and any polynomial expression can be expressed as a linear combination of the basis elements. For instance one can express the aforementioned polynomial as:

$$P(a) = \sum_{i=1}^{n} P^i(a) e_i, \qquad (A.12)$$

in which the $P^i(a)$ are polynomials in $a$ that take their values in $\mathbb{K}$.

When $V$ is also given a topology, which is usually assumed to make the algebra operations continuous, one can then speak of the convergence of a formal power series, as well; such an algebra $\mathfrak{A}$ is then called a *topological algebra*. Hence, for an associative topological algebra with unity we can define:

$$\exp(a) = \sum_{n=0}^{\infty} \frac{1}{n!} a^n. \qquad (A.13)$$

When the series converges, exp(a) is another element of $\mathfrak{A}$; when this series does not converge, it is to be interpreted as a formal power series.

In either event, if $\mathfrak{A}$ has a finite basis then one can express $A = \exp(a)$, when it converges, as a finite linear combination:

$$A = \sum_{i=1}^{n} A^i e_i. \qquad (A.14)$$

Each component $A^i$ will be defined by a power series in a variable that takes its values in $\mathbb{K}$. Furthermore, the nature of the power series will involve not only $a$, but also the structure constants of $\mathfrak{A}$, which relate to the way that successive powers of $a$ expand into components themselves.

In the example of $M(n; \mathbb{K})$, we have an associative algebra whose unity element is given by the identity matrix $I$. When the underlying vector space is given the product topology of $\mathbb{K}^{n^2}$, all expressions of the form exp($a$) will converge. Furthermore, since one has that:

$$\exp(\mathrm{Tr}(a)) = \det(\exp(a)) > 0 \qquad (A.15)$$

any matrix in the image of the map exp will be invertible. In particular, $\exp(0) = I$ and $\exp(-a) = [\exp(a)]^{-1}$. Hence, if $\mathfrak{A}$ is a sub-algebra of $M(n; \mathbb{K})$ then the image $A$ of $\mathfrak{A}$ under exp will be a subgroup of $GL(n; \mathbb{K})$. However, the map exp: $\mathfrak{A} \to A$ is not generally a homomorphism of the additive group on *A*, as it is when $\mathfrak{A} = \mathbb{R}$. Indeed, in order for one to have $\exp(a + b) = \exp(a)\exp(b)$ it is necessary and sufficient that the elements *a* and *b* must commute.

### A.5  Complexification

We shall need a general notion of the "complexification" of an algebra in order to make sense of the sequence of algebras that we shall be concerned with, namely $\mathbb{C}$, $\mathbb{H}$, and $\mathbb{H}^\mathbb{C}$. With that notion, one can then regard $\mathbb{C}$ as the complexification of $\mathbb{R}$ and $\mathbb{H}^\mathbb{C}$ as the complexification of $\mathbb{H}$, although the algebra $\mathbb{H}$ is *not* the complexification of the algebra $\mathbb{C}$, even though the complex vector space $\mathbb{C}^2 = \mathbb{R}^2 \oplus i\mathbb{R}^2$ can be regarded as the complexification of the real vector space $\mathbb{C} = \mathbb{R}^2$.

The usual way [**18**] of defining the *complexification* $V^\mathbb{C}$ of a real vector space *V* over the scalar field $\mathbb{K}$, which we always assume to contain $\mathbb{R}$ as a subfield, is to look at the vector space $V \otimes_\mathbb{R} \mathbb{C}$ of all *real*-linear maps from *V* to $\mathbb{C}$, which one then must regard as $\mathbb{R}^2 = \mathbb{R} \oplus i\mathbb{R}$. The real subspace of $V^\mathbb{C}$ then consists of all real-linear maps of *V* into the first summand of $\mathbb{R} \oplus i\mathbb{R}$, while the imaginary subspace *jV* consists of all real-linear maps of *V* into the second one. (We use the notation *j* for the imaginary vectors to distinguish them from imaginary numbers.) Hence, one can decompose $V^\mathbb{C}$ into $V \oplus jV = (V \otimes_\mathbb{R} \mathbb{R}) \oplus j(V \otimes_\mathbb{R} \mathbb{R})$. There is an obvious real-linear isomorphism $j: V \to jV$, $\mathbf{v} \mapsto j\mathbf{v}$. One can easily see that if the real dimension of *V* is *n* then the real dimension of $V^\mathbb{C}$ is $2n$.

In order to make $V^\mathbb{C}$ into a complex vector space one must define the multiplication of complex scalars time the elements of $V^\mathbb{C}$. This obtained by the use of the isomorphism that is defined by *j*:

$$(\alpha + i\beta)\mathbf{v} = \alpha\mathbf{v} + \beta j\mathbf{v}. \tag{A.16}$$

As particular examples, the complexification of $\mathbb{R}$ is clearly $\mathbb{C}$ and, similarly, any $\mathbb{R}^n$ complexifies to $\mathbb{C}^n$.

When the isomorphism *j* is combined with the reverse (but not inverse) isomorphism of *jV* with *V* that takes *j***v** to – **v**, we can define a real-linear isomorphism:

$$J: V \times V \to V \times V, (\mathbf{v}, \mathbf{w}) \mapsto (-\mathbf{w}, \mathbf{v}).$$

Addition of vectors in $V \times V$ is defined by the usual Cartesian product rule. One can define scalar multiplication by complex scalars on vectors $\mathbf{v} \in V \times V$ by way of:

$$(\alpha + i\beta)\mathbf{v} = \alpha\mathbf{v} + \beta J\mathbf{v}, \qquad \alpha, \beta \in \mathbb{R}. \tag{A.17}$$

Hence, $J$ makes $V \times V$ into a complex vector space, which we denote by $\mathbb{C}V$.

When $V$ is a real vector space, one can see that the association of $(\mathbf{v}, \mathbf{w})$ with $\mathbf{v} + j\mathbf{w}$ makes $\mathbb{C}V$ isomorphic to $V^{\mathbb{C}}$ as a complex vector space. Its complex dimension will then equal the real dimension of $V$.

However, if $V$ is assumed to be a complex vector space then the complex scalar multiplication on $\mathbb{C}V$ might differ from the action of $\mathbb{C}$ on $V$; that is, the action of $i$ takes $V$ to itself, but the action of $J$ takes make $JV$ a distinct space from $V$. Hence, the complex dimension of $\mathbb{C}V$ will be *twice* the complex dimension of $V$. In such a case, it might be better to refer to $\mathbb{C}V$ as the "external complexification" of $V$, so as not confuse it with $V^{\mathbb{C}}$, which will still be $V$.

The operator $J$ can be expressed in block matrix form:

$$J = \left[\begin{array}{c|c} 0 & -I \\ \hline I & 0 \end{array}\right], \tag{A.18}$$

and any element complex number $a + ib$ can be represented by the matrix:

$$aI + bJ = \left[\begin{array}{c|c} aI & -bI \\ \hline bI & aI \end{array}\right]. \tag{A.19}$$

This allows us to extend the action of the group $\mathbb{Z}_2 = \{I, J\}$ to an action of $U(1) \cong SO(2)$ on $\mathbb{C}V$ that one refers to as the action of *duality rotations*, by analogy with the corresponding operations on 2-forms on a four-dimensional vector space when that vector space of 2-forms is given a complex structure (see Delphenich [**19**]).

A duality rotation $R(\theta)$ through an angle $\theta$ is then represented by the rule:

$$\begin{aligned} R(\theta)(\mathbf{a} + j\mathbf{b}) &= (\cos\theta I + \sin\theta J)(\mathbf{a} + j\mathbf{b}) \\ &= (\cos\theta\,\mathbf{a} - \sin\theta\,\mathbf{b}) + j(\sin\theta\,\mathbf{a} + \cos\theta\,\mathbf{b}), \end{aligned} \tag{A.20}$$

which means that $R(\theta)$ can be represented by the block matrix:

$$R(\theta) = \cos\theta I + \sin\theta J = \left\{\begin{array}{c|c} \cos\theta I & -\sin\theta I \\ \hline \sin\theta I & \cos\theta I \end{array}\right\}. \tag{A.21}$$

This allows one to always define a smooth (indeed, analytic) circular loop that connects any **a** to its dual image *j***a**.

Another canonical linear isomorphism that is always associated with any vector space of the form $V \times V$, and which plays a recurring role in quantum wave equations, is the *transposition* isomorphism $T: V \times V \to V \times V$, which takes (**v**, **w**) to (**w**, **v**), and therefore has the block matrix form:

$$T = \begin{bmatrix} 0 & I \\ \hline I & 0 \end{bmatrix}. \tag{A.22}$$

When $V$ supports an associative algebra $\mathfrak{A}$ with unity $e$, in order to complexify the algebra, as well as the vector space, one must require that the algebra product $V^{\mathbb{C}} \times V^{\mathbb{C}} \to V^{\mathbb{C}}$, $(a, b) \mapsto ab$ be *complex* bilinear. Hence, one must have:

$$(\alpha + i\beta)ab = a(\alpha + i\beta)b = \alpha ab + j\beta ab, \tag{A.23a}$$

$$(a + jb)(c + jd) = (ac - bd) + j(ad + bc). \tag{A.23b}$$

As a consequence, one must have:

$$(ja)b = a(jb) = j(ab), \quad (ja)(jb) = -ab. \tag{A.24}$$

The complexification of an associative algebra is associative and the complexification of a commutative algebra is commutative. Furthermore, complexification will take a unity element to a unity element, and the complexification of the center of an algebra will be the center of the complexified algebra.

Suppose one chooses a basis $\{\mathbf{e}_i, i = 1, \ldots, n\}$ for the vector space $V$ over $\mathbb{K}$ and a set of structure constants $a_{ij}^k$ for an algebra $\mathfrak{A}$ over $V$ relative to that choice of basis elements. Then the complexification $\mathfrak{A}^{\mathbb{C}} = \mathfrak{A} \oplus J\mathfrak{A}$ of the algebra $\mathfrak{A}$ has a basis composed of $\{\mathbf{e}_i, \overline{\mathbf{e}}_i = j\mathbf{e}_i, i = 1, \ldots, n\}$.

One must then establish the extra structure constants by using (A.24):

$$\mathbf{e}_i \overline{\mathbf{e}}_j = \overline{\mathbf{e}}_i \mathbf{e}_j = j\mathbf{e}_i \mathbf{e}_j, \qquad \overline{\mathbf{e}}_i \overline{\mathbf{e}}_j = -\mathbf{e}_i \mathbf{e}_j . \tag{A.25}$$

This allows us to deduce that [5]:

$$a_{i\bar{j}}^{\bar{k}} = a_{\bar{i}\,j}^{\bar{k}} = -a_{\bar{i}\bar{j}}^{k} = a_{ij}^{k},  \tag{A.26}$$

while the remaining structure constants that must be introduced vanish.

If $\{e_a, a = 1, …, k\}$ is a set of generators for $\mathfrak{A}$ then it will also be a set of generators for $\mathfrak{A}^{\mathbb{C}}$.

### A.6  Representations

An *m-dimensional representation* of an *n*-dimensional algebra $\mathfrak{A}$ is a linear map $\mathcal{D}: \mathfrak{A} \to M(m, \mathbb{K})$, $a \mapsto \mathcal{D}(a)$ that preserves the multiplication:

$$\mathcal{D}(ab) = \mathcal{D}(a)\mathcal{D}(b) \tag{A.27}$$

for all $a, b \in \mathfrak{A}$. Hence, every element of $\mathfrak{A}$ will correspond to an $m \times m$ matrix with entries in $\mathbb{K}$ and the multiplication of elements in $\mathfrak{A}$ will correspond to the multiplication of matrices in $M(m, \mathbb{K})$.

The representation $\mathcal{D}$ is *faithful* if it is a linear injection. In such a case, the algebra over the image of $\mathfrak{A}$ under $\mathcal{D}$ will be isomorphic to the algebra $\mathfrak{A}$.

If $\{\mathbf{e}_i, i = 1, …, k\}$ is a set of generators for $\mathfrak{A}$ ($k \leq n$) then it is sufficient to know the matrices $\mathcal{D}(\mathbf{e}_i)$ in order to determine the images of all products $ab$ under the representation $\mathcal{D}$.

If $a = a^i \mathbf{e}_i$ and $b = b^j \mathbf{e}_j$ and the structure constants for $\mathfrak{A}$ for this basis are $a_{ij}^k$ then one has:

$$\mathcal{D}(ab) = a_{ij}^k a^i b^j \mathcal{D}(\mathbf{e}_k). \tag{A.28}$$

If $\{e_a, a = 1, …, m^2\}$ is a basis for $M(m, \mathbb{K})$ then one can express the representation $\mathcal{D}$ in terms of its matrix $\mathcal{D}_a^i$ with respect to the choice of bases on $\mathfrak{A}$ and $M(m, \mathbb{K})$. By definition:

$$e_a = \mathcal{D}_a^i \mathbf{e}_i. \tag{A.29}$$

---

[5] Here, we are using the "kernel-index" notation for the structure constants, as preferred by Schouten and his followers. That is, the kernel in this case is the letter *a*, whereas the overbar on the indices denotes indices that refer to basis elements that have overbars.

Generally, in physics applications when the basis elements are matrices themselves one usually adds extra indices to $\mathcal{D}_a^i$ for the rows and columns.

The particular representations that we shall be concerned with are the defining representation, which only makes sense when $\mathfrak{A}$ is a subalgebra of $M(n; \mathbb{K})$ to begin with, and the *left* and *right multiplication* representations. That is, if $\mathfrak{A}$ is an algebra over a vector space $V$ then each $a \in \mathfrak{A}$ defines a left multiplication operator $L_a: \mathfrak{A} \to \mathfrak{A}$, $x \mapsto ax$ and a right multiplication operator $R_a: \mathfrak{A} \to \mathfrak{A}$, $x \mapsto xa$. These, in turn, define maps $L: \mathfrak{A} \to \mathrm{End}(\mathfrak{A})$, $a \mapsto L_a$, and $R: \mathfrak{A} \to \mathrm{End}(\mathfrak{A})$, $a \mapsto R_a$, which are easily seen to define representations of $\mathfrak{A}$ in the algebra $\mathrm{End}(\mathfrak{A})$ of all linear maps of $\mathfrak{A}$ to itself. Choosing a basis for $\mathfrak{A}$ then defines an isomorphism of $\mathrm{End}(\mathfrak{A})$ with $M(n; \mathbb{K})$.

It is easy to see that if A is a real algebra that is represented in a real vector space $V$ by way of $\mathcal{D}: \mathfrak{A} \to M(V)$ then there is a canonical extension of $\mathcal{D}$ to a representation of $\mathcal{D}$: $\mathfrak{A}^{\mathbb{C}} \to M(V^{\mathbb{C}})$, $\mathcal{D}(a + jb) \mapsto \mathcal{D}(a) + j\mathcal{D}(b)$.